\documentclass[12pt,preprint]{aastex}

\usepackage{natbib}
\usepackage{graphics}
\usepackage{epsfig}
\usepackage{amssymb}
\usepackage{rotating}
\usepackage{lscape}
\usepackage{longtable}

\shorttitle{STELLAR ROTATION IN M34}
\shortauthors{Meibom et al.}

\begin{document}

\title{THE COLOR-PERIOD DIAGRAM AND STELLAR ROTATIONAL EVOLUTION\\~
 - NEW ROTATION PERIOD MEASUREMENTS IN THE OPEN CLUSTER M34\altaffilmark{1}}

\author{S{\o}ren Meibom\altaffilmark{2,3}}
\affil{Harvard-Smithsonian Center for Astrophysics, Cambridge, MA, 02138, USA}

\author{Robert D. Mathieu}
\affil{Department of Astronomy, University of Wisconsin - Madison, Madison,
WI, 53706, USA}

\author{Keivan G. Stassun\altaffilmark{4}}
\affil{Physics and Astronomy Department, Vanderbilt University, Nashville, TN, 32735, USA}

\author{Paul Liebesny}
\affil{Department of Physics Math \& Science, Emory University, Atlanta, GA, 30322, USA}

\author{Steven H. Saar}
\affil{Harvard-Smithsonian Center for Astrophysics, Cambridge, MA, 02138, USA}

\altaffiltext{1}{WIYN Open Cluster Study XLV}
\altaffiltext{2}{{\it smeibom@cfa.harvard.edu}}
\altaffiltext{3}{Visiting Astronomer, Kitt Peak National Observatory,
National Optical Astronomy Observatory, which is operated by the
Association of Universities for Research in Astronomy, Inc. (AURA)
under cooperative agreement with the National Science Foundation.}
\altaffiltext{4}{Also Physics Department, Fisk University, Nashville, TN, 37208}


\begin{abstract} \label{abs}

We present the results of a 5-month photometric time-series survey for
stellar rotation periods combined with a 4-year radial-velocity survey for
membership and binarity in the 220\,Myr open cluster M34. We report surface
rotation periods for 120 stars, 83 of which are kinematic and photometric
late-type cluster members. A comparison to previous work serves to illustrate
the importance of high cadence long baseline photometric observations and
membership information. The new M34 periods are less biased against slow
rotation and cleaned for non-members. The rotation periods of the cluster
members span over more than an order of magnitude from 0.5 day up to 11.5 days,
and trace two distinct rotational sequences - fast ($C$) and moderate-to-slow
($I$) - in the color-period diagram. The sequences represent two different
states (fast and slow) in the rotational evolution of the late-type cluster
members. We use the color-period diagrams for M34 and for younger and older
clusters to estimate the timescale for the transition from the $C$ to the
$I$ sequence and find $\la$150\,Myr, $\sim$150-300\,Myr, and $\sim$300-600\,Myr
for G, early-mid K, and late K dwarfs, respectively. The small number of stars
in the gap between $C$ and $I$ suggest a quick transition. We estimate a lower
limit on the maximum spin-down rate ($dP/dt$) during this transition to
be $\sim$0.06\,days/Myr and $\sim$0.08\,days/Myr for early and late K dwarfs,
respectively. We compare the $I$ sequence rotation periods in M34 and
the Hyades for G and K dwarfs and find that K dwarfs spin down slower than
the Skumanich $\sqrt{t}$ rate. We determine a gyrochronology age of 240\,Myr
for M34. The gyro-age has a small formal uncertainty of 2\% which reflects
the tight $I$ sequence in the M34 color-period diagram. We measure the effect
of cluster age uncertainties on the gyrochronology age for M34 and find the
resulting error on the gyro-age to be consistent with the $\sim15\%$
error estimate for the technique in general. We use the M34 $I$ sequence
to redetermine the coefficients in the expression for rotational dependence
on color used in gyrochronology. Finally, we propose that stability in the
phase, shape, and amplitude of the photometric variability for the 120
rotators over the $\sim$5 month duration of our survey, is due to spot
generation at active stellar longitudes.

\end{abstract}

\keywords{open clusters and associations:general - stars:late-type
 - stars:rotation - stars:ages - stars:spots}


\section{INTRODUCTION}
\label{intro}

The angular momentum of a late-type star is not conserved and is
depleted at varying rates over its lifetime. The changes with time of this
fundamental stellar property is an integral part of the formation and
evolution of the Sun and stars similar to it. Our understanding of the
rate and timing of the depletion of stellar angular momentum is tied to
our understanding of the physical mechanisms responsible for redistribution
of angular momentum inside of stars, and for the generation and evolution
of the magnetic fields by which angular momentum is lost externally.
Observing the rotational evolution for late-type stars of different
masses, and thus different internal structures, is a critical step toward
identifying which physical processes are at work and when.

Open star clusters present coeval and homogeneous populations of stars
over a range of masses and with well known ages. They are therefore ideal
systems for studying stellar rotation as a function of mass and age.
Calibrated relationships, based on observations in clusters, allow rotation
to be used as an indicator of stellar age \citep[gyrochronology, see e.g.][]
{barnes03a,barnes07,mh08,soderblom10}.
An important early example is the study by \citet{skumanich72} of solar-type
stars in the Pleiades and Hyades from which the relation between stellar
projected rotation velocity and age ($v\sin i \propto t^{-0.5}$) derive.
Subsequent observations of $v\sin i$ for late-type dwarfs in young
($\la600\,Myr$) open clusters revealed bimodality and large dispersions
in rotation rates, but no clear relationship to stellar mass
\citep[e.g.][]{sjw83,ssh+93,shs+84,shb+85,shj89,sh87,tsp+00,sjf01,tpj+02}.
In the Hyades \citep[625\,Myr;][]{pbl+98}, \citet{rtl+87} found that
the rotation of even the most rapidly spinning F, G, and K (FGK) dwarfs
has converged toward a mass-dependent rate. Such observations combined
with rotational data for late-type stars in the pre main-sequence (PMS)
phase \citep[see][for a review]{hem+07}, have shown that their
early angular momentum evolution is highly dynamical and that there are
gaps in our understanding of it.

Models of the evolution of angular momentum have developed in parallel
with the observations and have become increasingly sophisticated. The
angular momentum loss rate for a late-type star is expressed in terms
of its spin velocity, mass, radius, and mass loss rate through a magnetized
stellar wind \citep[e.g.][]{kawaler88,pks+89,bfa97,spt00,bsp01,dpt+10}.
The loss rates predicted by \citet{wd67} and subsequent theoretical
studies are consistent with the observed Skumanich relation under certain
assumptions about the production (a linear dynamo) and geometry of the
stellar magnetic field. However, to account for the observed evolution
from the PMS phase to the Sun, current models have come to include additional
elements such as ``saturation'' of the wind-loss at a mass-dependent
threshold rotation velocity \citep[e.g.][]{mb91,cdp95,bs96,kpb+97,spt00,tpt02},
magnetic braking of stellar rotation by a circumstellar disk during the
PMS phase \citep[``disk-locking'', e.g.][]
{konigl91,sno+94,kpb+97,bsp01,spt00,dpt+10} and decoupling followed by
re-coupling on a mass-dependent timescale of the star's radiative core
and convective envelope \citep[e.g.][]{es81,mb91,jc93,dpt+10}. The
increasing complexity of the models stresses the importance of obtaining
and analyzing new and better observational data, especially at ages
between the Pleiades ($\sim$125\,Myr) and the Hyades ($\sim$625\,Myr).

Presently, measurements of the surface rotation velocity or period provide
the best empirical constraints on stellar angular momentum models. Early
studies in open clusters measured primarily $v \sin i$, requiring only one 
measurement of the Doppler broadening of spectral lines to produce a result.
This technique is useful for stars that rotate faster than a threshold
velocity set by the spectral resolution, but the result is ambiguous due
to the unknown stellar radius ($R$) and spin-axis inclination ($i$).
Accordingly, recent work has focused on direct measurements of the stellar
rotation period from brightness modulations by star spots \citep[e.g.][]
{psd+95,ktp+98,barnes03a,iha+07,mms09,cdh+09,hgp+09,hbk+10}. The rotation
period is independent of $i$ and $R$ and only mildly affected by latitudinal
differential rotation. Its detection is limited to stars with periodic
brightness variations with amplitudes above a threshold set by the
photometric precision and requires access to uninterrupted telescope time
with a duration of at least twice the number of nights of the longest
period one wants to measure. Nevertheless, dedicated time-series photometric
surveys of cool stars in young open clusters have begun to reveal mass- and
time-dependencies on stellar rotation not previously discernible in the
$v \sin i$ data \citep[e.g.][]{barnes03a,hgp+09} - and in particular when
combined with information about cluster membership to produce the cleanest
samples \citep{mms09,cdh+09,hbk+10}.

The emergence from these studies of distinct groups of coeval fast and
slowly rotating stars, forming tight ``rotational sequences'' in the
color vs. rotation period plane, have opened a new window on the angular
momentum evolution of late-type stars. Indeed, the observed time- and
mass-dependent changes in the distributions of stellar rotation period
vs. color, promise to make the {\it color-period diagram} an important
testing ground for future models of angular momentum evolution - much
like the color-magnitude diagram was and still is for models of stellar
evolution.

This paper joins a series of studies to employ the color-period diagram
in order to empirically describe the relationship between stellar surface
rotation, age, and mass for late-type stars. Our earlier study
\citep[hereinafter Paper I;][]{mms09} targeted the $\sim$180\,Myr
open cluster M35. Here we present the results of an extensive multi-month
time-series photometric survey for rotation periods and a multi-year
spectroscopic survey for membership and binarity in the field of the open
cluster M34 (NGC\,1039).
M34 is a northern hemisphere cluster located $\sim$470 pc \citep{jp96}
toward the galactic anti-center ($\alpha_{2000} = 2^{h}~42^{m}$,
$\delta_{2000} = 42\degr~45\arcmin$; $l = 143\fdg7$, $b = -15\fdg6$)
with an estimated reddening of $E_{B-V} = 0.07$ \citep{ccp79}. Recent
age estimates derived from fitting stellar evolution isochrones to the
cluster color-magnitude diagrams range from 177\,Myr \citep{mmm93} to
251\,Myr \citep{is93}. \citet{jbm+10} derived a mean gyrochronology age of
$198 \pm 9$\,Myr. We adopt a fiducial (average) stellar evolution age
of 220\,Myr for this paper. The metallicity of M34 is found to be close
to solar \citep[Fe/H $\sim0.07 \pm 0.04$;][]{skf+03,steinhauer10}.

Section~\ref{obs} describes the photometric and spectroscopic observations,
data-reduction, and determination of rotation periods and cluster membership.
Section~\ref{results} presents our results and comparisons to results from
prior rotational studies in M34. Section~\ref{analysis} offers analysis and
discussion of the M34 color-period diagram in the context of early angular
momentum evolution and of gyrochronology. Summary and conclusions are given
in Section~\ref{conclusions}.


\section{OBSERVATIONS, DATA-REDUCTION, AND ROTATION PERIOD DETERMINATION}
\label{obs}

The photometric and spectroscopic observations presented in this paper
for stars in M34 were taken in parallel with the observations for Paper I.
We have reduced and analyzed the M34 data using the same procedures and
methods. Thus, we provide here only basic information about the observations,
data-reduction, and determination of rotation periods and cluster membership,
and refer to \citet{mms09} for a more detailed description.

\subsection{Time-series photometry}
\label{tsp}

\subsubsection{Observations}

We obtained time-series photometry over a time-span of 143 days for a
$40\arcmin \times 40\arcmin$ region centered on the open cluster M34.
The photometric data were obtained in the Johnson V-band with the WIYN
0.9m telescope
\footnote{The 0.9m telescope is operated by WIYN Inc. on behalf of a
Consortium of ten partner Universities and Organizations (see
http://www.noao.edu/0.9m/general.html)}
on Kitt Peak equipped with a $20\arcmin \times 20\arcmin$ field-of-view
$2k \times 2k$ CCD camera.

The dataset presented is composed of images from high-frequency
(approximately once per hour for 5-6 hours per night) time-series
photometric observations over 16 full nights from 2-17 December 2002,
complemented with a queue-scheduled observing program over 143 nights
from 22 October 2002 to 11 March 2003, obtaining one image per night
interrupted only by bad weather and scheduled instrument changes. The
result is a database of differential V-band light curves for 5,656 stars
with $12 \la V \la 20.8$.

The sampling frequency of the December 2002 observations allows
us to detect photometric variability with periods ranging from
less than a day to about 10 days. The long time-span of the
queue-scheduled observations provide data suitable for detecting
periodic variability of up to $\sim$ 75 days, and for testing
the near-term stability of photometric variations. From this
database we derive rotation periods for 120 stars.

\begin{figure}[ht!]
\epsscale{1.0}
\plotone{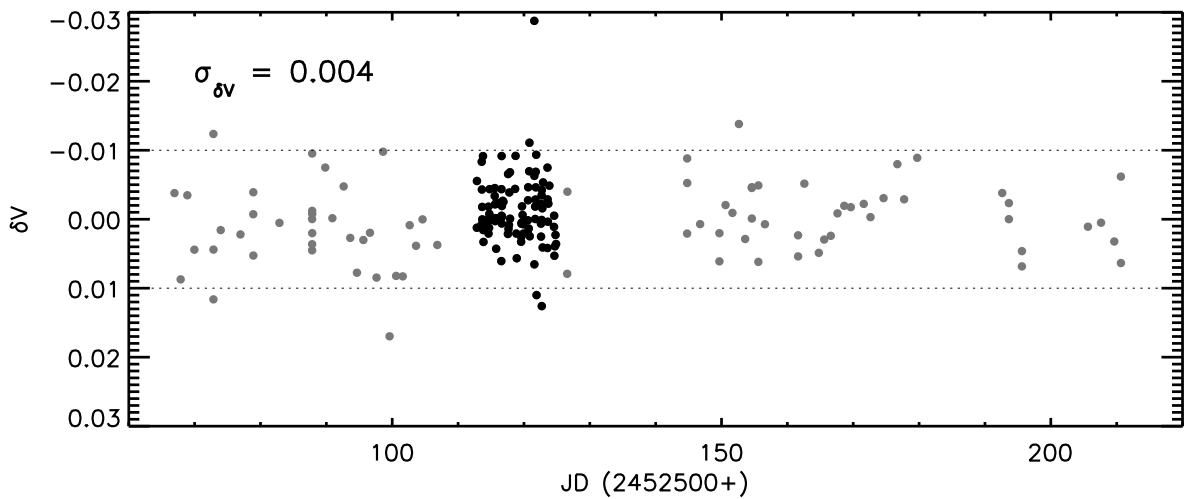}
\caption{Sample time-series data from from our photometric database for
a non-variable $V \simeq 13\fm5$ star. Black symbols represent
measurements from the high-frequency December 2002 observing run and grey
symbols represent the low-frequency queue-scheduled observations. The data
span a total of 143 days. The star was observed in 178 images. The standard
deviation ($\sigma_{\delta V}$) of the 178 measurements is $\sim 0\fm004$,
representative of our best photometric precision. The horizontal dotted
lines denote $\delta V = \pm 0\fm01$.
\label{sampling}}
\end{figure}

Figure~\ref{sampling} displays the time-series data from both observational
programs for a photometrically non-variable star. Black symbols represent
the high-frequency observations and grey symbols represent the queue-scheduled
observations.

\subsubsection{Basic reductions, photometry, and light curves}

Basic reductions of our CCD frames, identification of stellar sources,
and computations of equatorial coordinates\footnote{We used data from
the STScI Digitized Sky Survey; The Digitized Sky Surveys were produced
at the Space Telescope Science Institute under U.S. Government grant
NAG W-2166.} were done using standard IRAF packages. Instrumental
magnitudes were determined from Point Spread Function (PSF) photometry
using the IRAF DAOPHOT package. The analytical PSF and a residual look-up
table were derived for each frame based on $\sim$30 evenly distributed
isolated stars. The number of measurements in the light curves 
vary because stars near the edges of individual frames may be
missed due to telescope pointing errors, while bright stars near the
CCD saturation limit and faint stars near the detection threshold may
be excluded on some frames because of variations in seeing, sky brightness,
and sky transparency. To ensure our ability to perform reliable
time-series analysis on stars in our database, we have eliminated
stars that appear on fewer than 50 frames.

\begin{figure}[ht!]
\epsscale{1.0}
\plotone{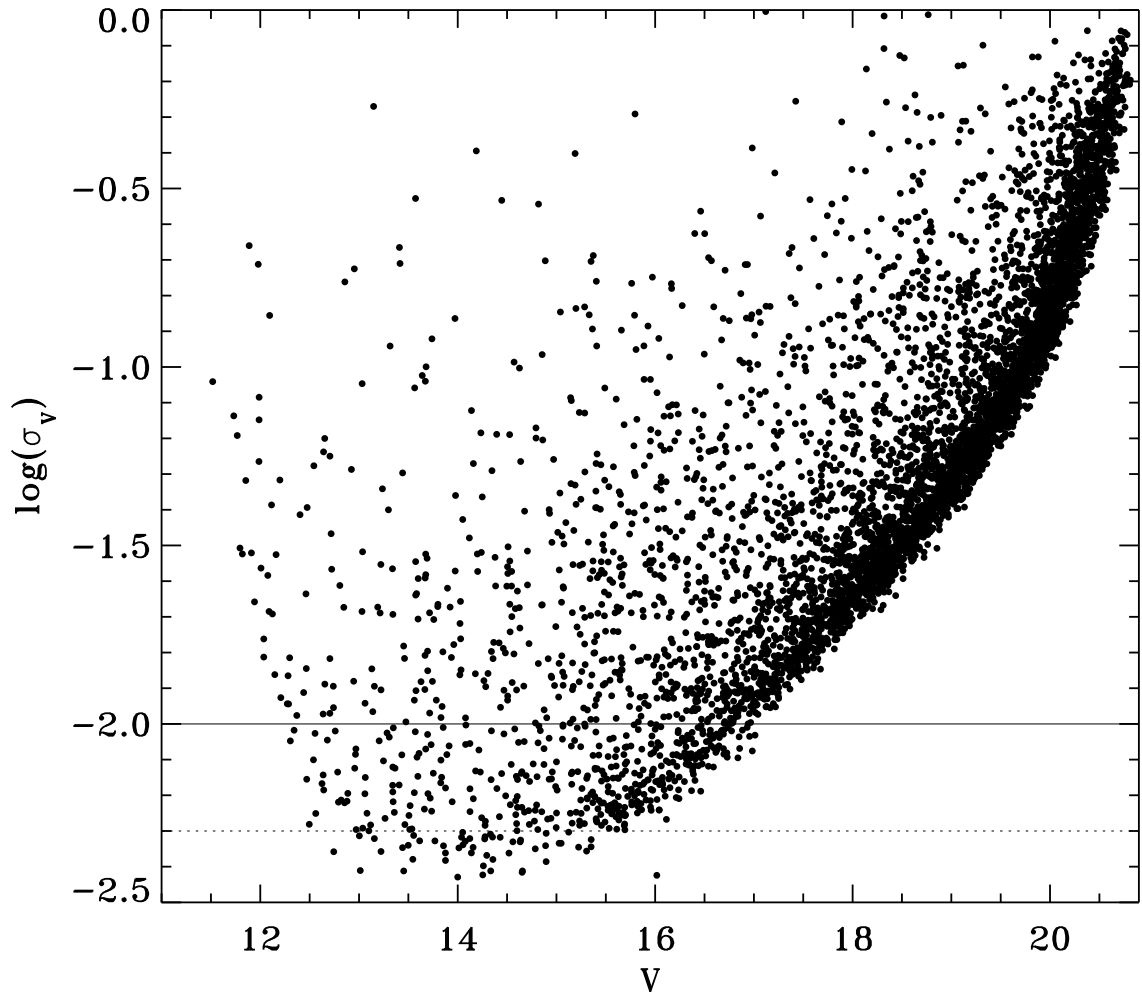}
\caption{The logarithm of the standard deviation of all instrumental
magnitudes as a function of apparent V magnitude for 5,656 stars in the field
of M34. The number of measurements for each star range from 50 to 178.
The solid and dashed horizontal lines represent $\sigma_V$ of 0.01 (1\%)
and 0.005 (0.5\%), respectively.
\label{m0_sigst}}
\end{figure}

We applied the \citet{honeycutt92} algorithm as implemented in \citet
{smm+99,smv+01} for differential CCD photometry to our raw light curves
to remove non-stellar frame-to-frame photometric variations.
Figure~\ref{m0_sigst} shows the standard deviation of the photometric
measurements as a function of the apparent V magnitude for each star
in the field of M34. The relative photometric precision is $\sim$0.5\%
for stars with $12.5 \la V \la 15.5$.

\subsubsection{Photometric period detection}

We employed the \citet{scargle82} periodogram analysis to detect
periodic variability in the light curves because of its ability to
handle unevenly sampled data. We searched 5000 frequencies corresponding
to periods between 0.1 day and 90 days. The lower search limit was set
at a period ensuring critical sampling based on the Nyquist critical
frequency for our high-frequency data ($f_{c} = 1/(2 \delta t)$, where
$\delta t$ is the sampling interval of $\sim$1 hour). The upper limit
was set at 90 days because a star with a 90-day period would complete
about 1.5 cycle over the 143 nights of the survey.

A false alarm probability (FAP), the probability that a signal
detected at a certain power level can be produced by statistical
fluctuation, was calculated as the measure of confidence in a
detected period. The FAP for each star is based on a two-dispersion
Monte Carlo calculation generating 100 synthetic light curves.
The maximum power of the resulting 100 periodograms was adopted
as the level of 1\% FAP, and used as the initial threshold for
detecting significant photometric variability. For all stars that
met the FAP criterion we examined - by eye - the periodogram and raw
and phased light curves. We report stellar rotation periods for 120
stars in our database (see Table 1 in Appendix B).


\subsection{The Spectroscopic Survey}

We observed M34 from 2001 to 2004 as part of the WIYN
\footnote{The WIYN Observatories are joint facilities of the University
of Wisconsin-Madison, Indiana University, Yale University, and the National
Optical Astronomy Observatories.}
Open Cluster Study (WOCS; \citet{mathieu00}). A total of 3320 spectra has
been obtained of 437 FGK stars within a 1-degree field centered
on the cluster. The selection of spectroscopic target stars was based on
photometric membership data (\citet{deliyannis10}, see Section~\ref{phot}).
In the M34 $B-V$ vs. $V$ color-magnitude diagram (CMD), stars on or less
than $1\fm0$ above and less than $0\fm5$ below the cluster main
sequence were selected between $V$ of 12 ($\sim1.4~M_{\odot}$) and 17
($\sim0.6~M_{\odot}$). All spectroscopic data were obtained using the
WIYN 3.5m telescope equipped with a multi-object fiber optic positioner
(Hydra) feeding a bench mounted spectrograph.  Observations were done
at central wavelengths of 5130\AA\ or 6385\AA\ with a wavelength range
of $\sim$200\AA\, including many narrow absorption lines as well as the
Mg B triplet. Radial velocities with a precision of $\la 0.5~km~s^{-1}$
\citep{gmh+08} were derived from the spectra via cross-correlation with
a high $S/N$ sky spectrum.

Of the 120 stars with rotation periods presented in this study, 74 have
one or more radial-velocity measurements (the remainder being below the
faint limit of the spectroscopic survey or photometric non-members). The
radial-velocity cluster membership probability of each star is calculated
using the formalism by \citet{vkp58} and the stellar mean radial velocity.
We have adopted $P_{RV}=40\%$ as the threshold for assigning radial-velocity
cluster membership. A separate paper describing the spectroscopic survey
of M34 and presenting the resulting membership information will follow.


\subsection{The M34 Color-Magnitude Diagram and Cluster Membership}
\label{phot}

\begin{figure}[ht!]
\epsscale{1.0}
\plotone{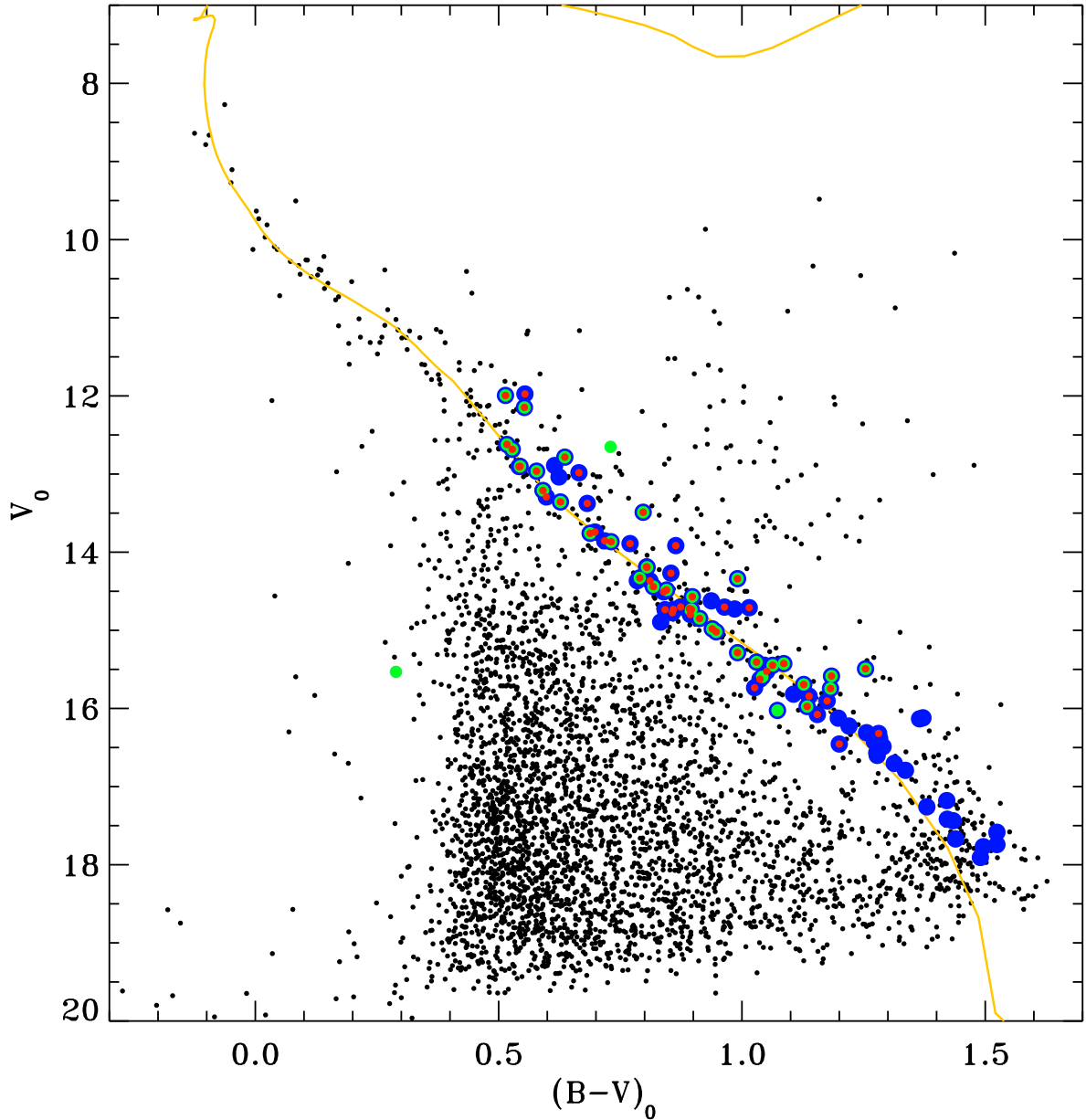}
\caption{Color-magnitude diagram based on photometry by \citet{deliyannis10}.
Stars marked in blue, green, and red, represent stars with measured rotation
periods that are photometric, proper-motion, or radial-velocity cluster
members, respectively. The 200\,Myr Yale isochrone used to convert stellar
color to stellar mass for the M34 members is shown in orange/yellow.
\label{cmd}}
\end{figure}

Figure~\ref{cmd} shows the $B-V$ vs. $V$ color-magnitude diagram (CMD)
for M34. The photometry was kindly provided by \citet{deliyannis10}
who obtained BVRI data over a $\sim40\arcmin \times 40\arcmin$ field
on the cluster using the WIYN 0.9m telescope. 
We highlight only those of the 120 rotators that have been found to be either
photometric, proper-motion \citep{jp96}, or radial-velocity
cluster members. Stars selected as photometric members are shown
in blue. Stars that are proper-motion members are overplotted in green. 
Radial-velocity members are marked in red. The fraction of late-F and
G dwarf members in M34 with measured rotation periods is 22\%, and the
fraction of K dwarf members in M34 with measured rotation periods is
23\%. We thus find no difference in the detection rate as a function
of stellar color (mass) or brightness in the range of interest to this
study.

Of the 92 photometric members 9 are kinematic non-members bringing
the sample of cluster members with rotation periods to 83 stars. These
83 stars constitute our study sample for M34 in this paper.

\section{RESULTS}
\label{results}

\subsection{Rotation periods}

\begin{figure}[ht!]
\epsscale{1.0}
\plotone{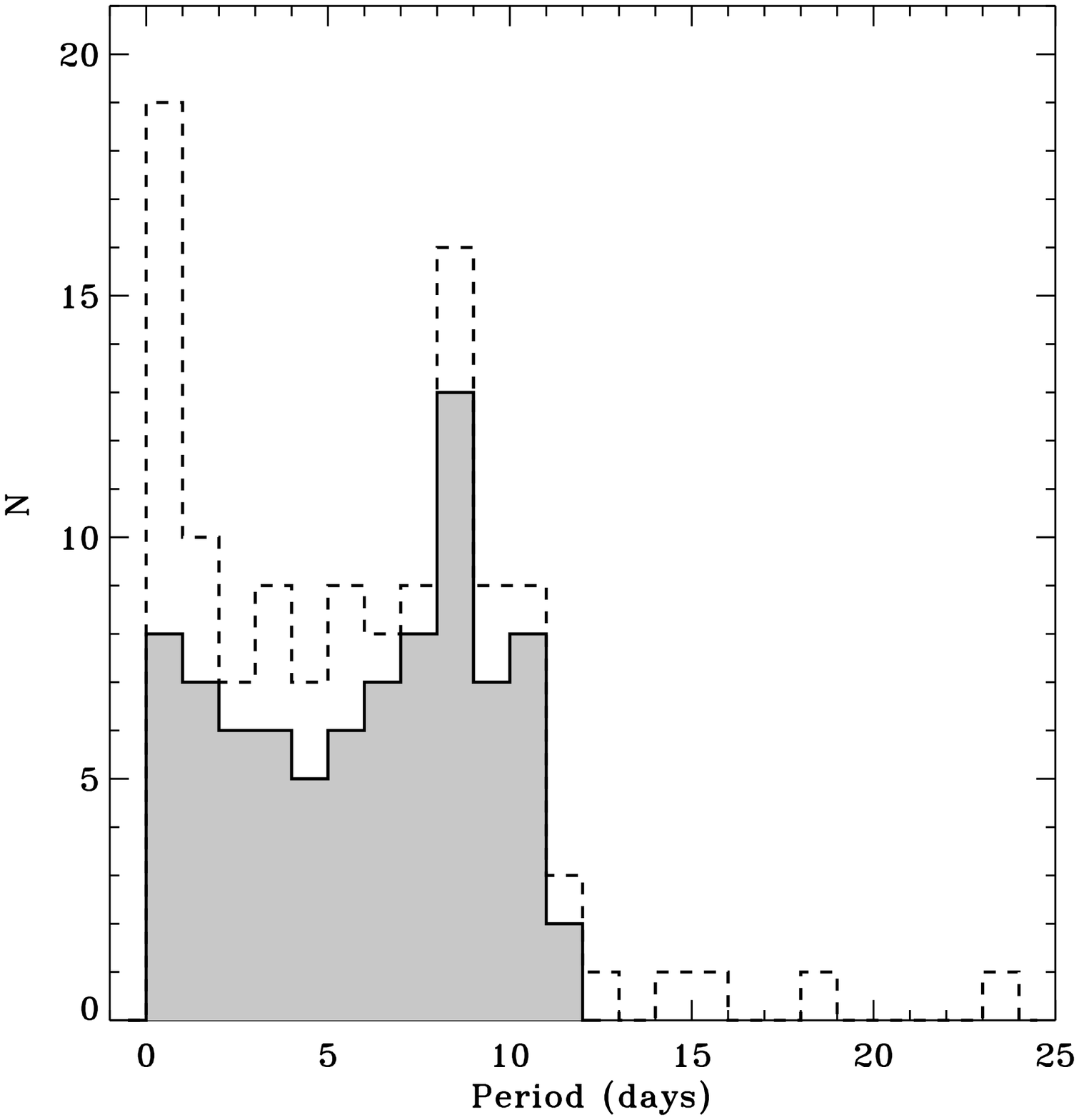}
\caption{The distribution of all 120 surface rotation periods determined
in this study (dashed line) and of the 83 periods for members of M34 (grey
histogram).
\label{pdistr}}
\end{figure}

Figure~\ref{pdistr} shows the distribution of all 120 rotation periods
determined in this study. The distribution spans over more than an order
of magnitude from from $\sim$0.1 days to $\sim$23 days, showing that we are
capable of detecting rotation periods down to the pseudo-Nyquist period-limit
of about 2 hours ($\sim0.08$ days) resulting from our sampling cadence
of about $1~hr^{-1}$ in December 2002. The shortest rotation period for
a cluster members is 0.49 days, suggesting that the lower limit of 0.1
days for our period search was set appropriately for the stars in M34.
The long-period tail of the period distribution for all 120 rotators
falls off at $\sim$12-13 days, but 4 stars have longer periods.
The longest rotation period among cluster members is 11.5 days, while
rotation periods of $\sim$14, 15, 18, and 23 days are found among field
stars. This may suggest that we observe a physical upper limit on the
rotation period distribution among FGK dwarfs in M34. It is possible
that we are not detecting the photometric variability of members with
even slower rotation if the frequency
and size of spots on such stars is insufficient for detection with the
photometric precision of our data \citep[e.g.][]{sab+04,svf07}. Constraining
the rotation for such slowly rotating stars will require either higher
photometric precision \citep[see e.g.][]{hbn+10,ibb+10} or high-resolution
($R \ga 50,000$) spectroscopic observations to measure projected rotation
velocities.

\subsection{The M34 color-period diagram}

\begin{figure}[ht!]
\epsscale{1.0}
\plotone{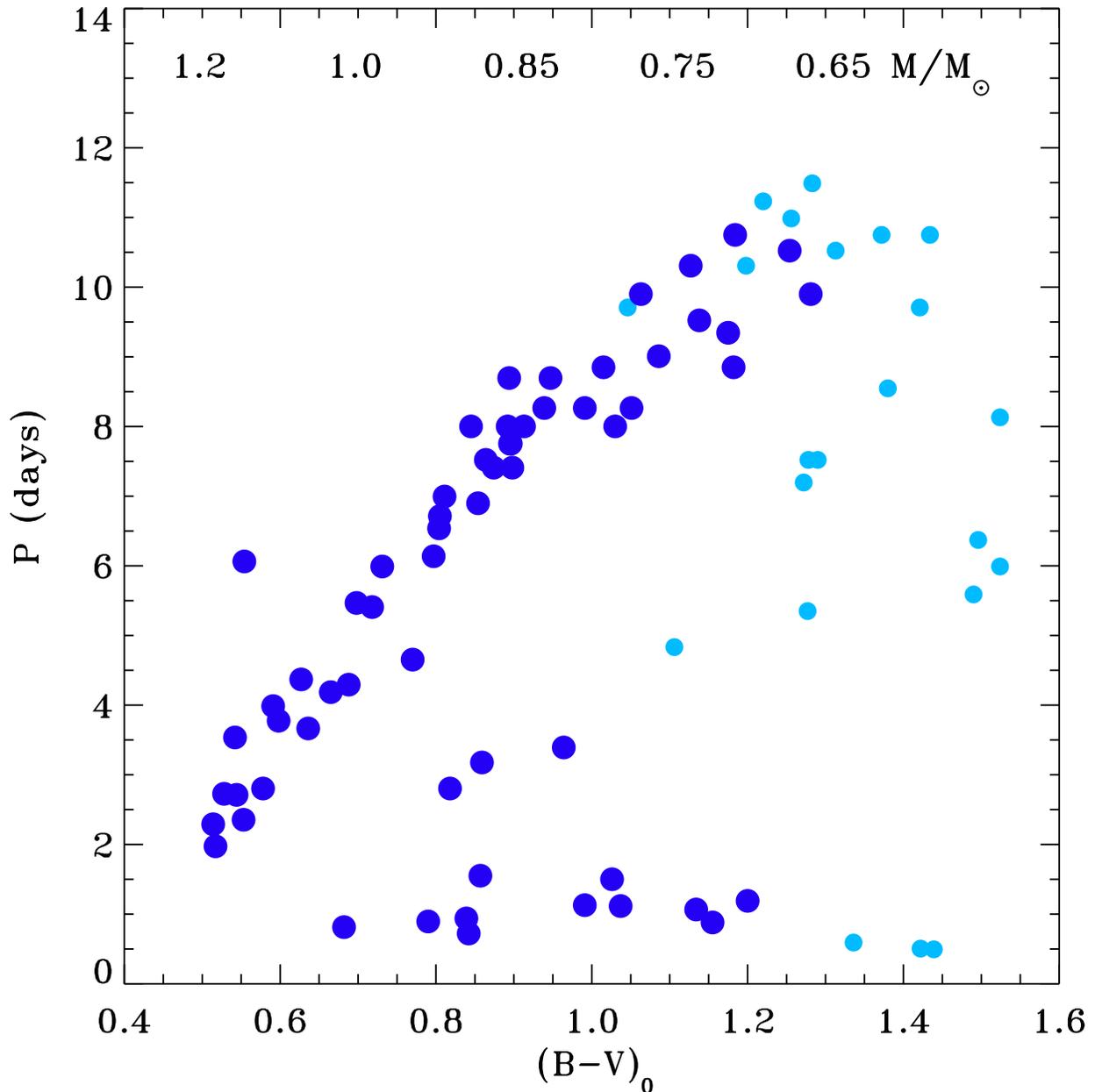}
\caption{The distribution of surface rotation periods with B-V color
index for 83 members of M34. Dark blue symbols represent stars that are
both photometric and kinematic cluster members. Light blue symbols are
used for stars that are photometric members only. The upper x-axis gives
a stellar mass estimate for dwarfs of corresponding color.
\label{pbv}}
\end{figure}

We focus now on the 83 members of M34 with measured rotation periods.
In Figure~\ref{pbv} we display their rotation periods against their
$B-V$ color indices. Dark blue symbols represent stars that are both
photometric and kinematic (radial-velocity and/or proper-motion) members
of M34. Light blue symbols are used for stars that are photometric members
only. The color indices derive from the multi-band photometry by
\citet[][Section~\ref{phot}]{deliyannis10} corrected for a reddening
of $E_{B-V}=0.07$. The corresponding stellar masses are estimated
from a 200\,Myr Yale isochrone \citep[][see Figure~\ref{cmd}]{ykd03}
and marked on the upper x-axis. As in Paper I, we use the dereddened
color index as the primary independent variable because the
corresponding stellar mass is model dependent and includes additional
errors and uncertainties in the color-mass transformation.

\placefigure{pbv}

\notetoeditor{We would like  Figure~\ref{pbv} in color and large - spanning
both column in width.}

M34 is less rich and closer than M35 (Paper I) and in a less crowded stellar
field. Therefore, on average, more radial-velocity measurements were obtained
on any given candidate cluster member and to a fainter magnitude limit. Stars
with variable radial velocities were observed more frequently than velocity
stable stars. The median number of velocity measurements for radial-velocity
members is 10, with a minimum and a maximum of 3 and 18 velocities,
respectively. Accordingly, {\it the M34 color-period diagram represent an
exceptionally ``clean'' sample of coeval single and close binary FGK dwarfs},
showing the preferred stellar rotation periods
as a function of color for 220\,Myr late-type stars. The 83 members lay
primarily along two rotational sequences. One sequence displays a narrow
diagonal band of stars whose periods are increasing with increasing color
(decreasing mass). The second sequence consists of rapidly rotating ($P_{rot}
\la 2$ days) stars. A subset of stars are found distributed between these
two sequences, and one late-F single member has a rotation period about twice
that of similar cluster stars placing it above the diagonal sequence in
Figure~\ref{pbv}.

\begin{figure}[ht!]
\epsscale{1.0}
\plotone{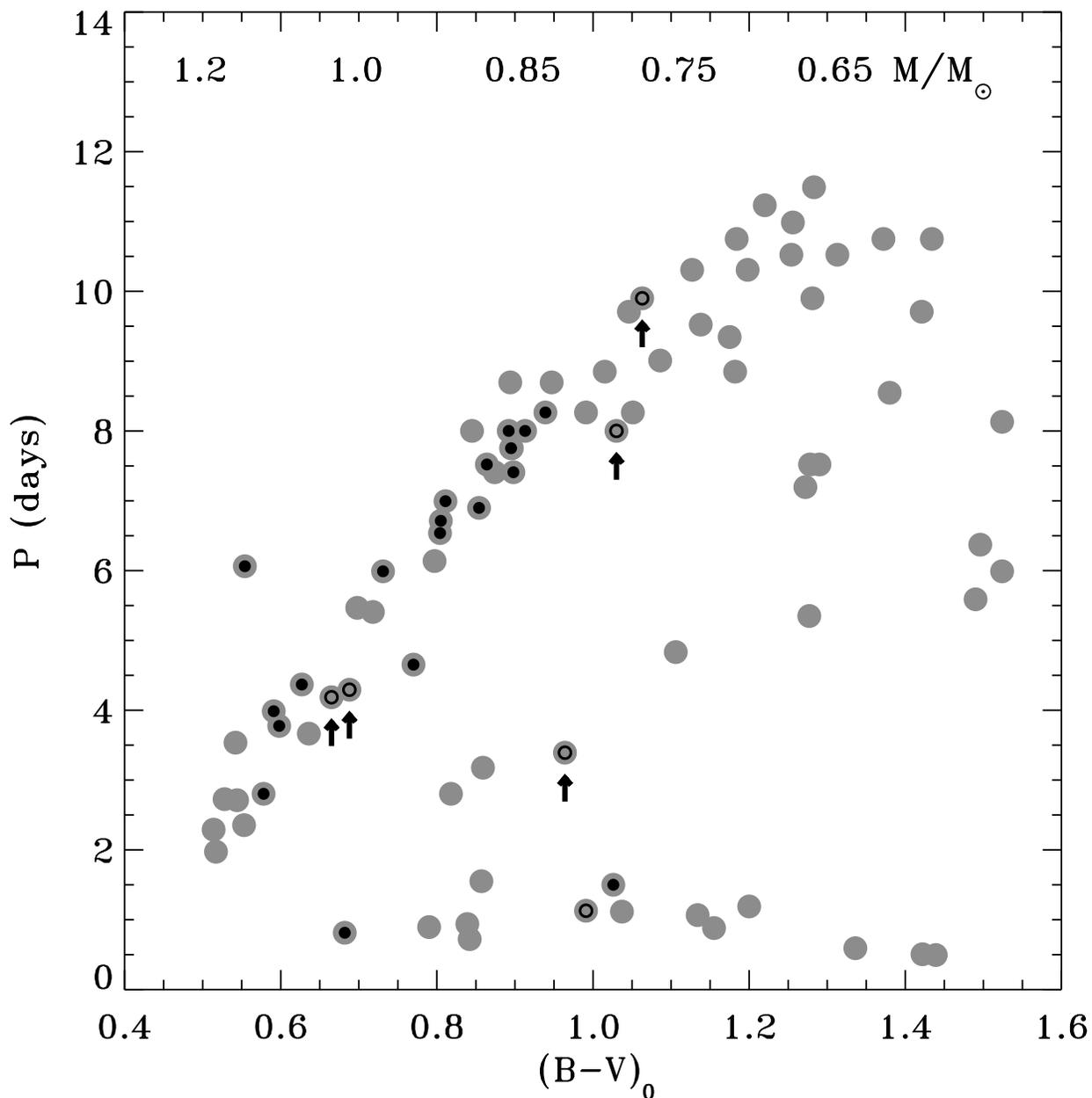}
\caption{The distribution of stellar rotation periods with B-V color
index for 83 members of M34. Nineteen single stars are marked with
additional filled black circles, and 6 close spectroscopic binaries
are marked with additional open circles. Five binary stars have
spectroscopic orbits and are marked with arrows.
\label{pbv_sm_bm}}
\end{figure}

Our spectroscopic survey allow us to identify single and close binary stars
among the M34 members. Figure~\ref{pbv_sm_bm} shows the location in the
color-period diagram of 19 single and 6 close binary members. The single
stars are marked with additional filled black circles. These stars have 5
or more radial-velocity measurements with a standard deviation of less
than $0.5~km~s^{-1}$ (our radial-velocity measurement precision). Stars
identified as close binaries are marked with additional open circles.
These stars have between 14 and 18 radial-velocity measurements with a
standard deviation above $2.5~km~s^{-1}$ ($5\sigma$). The remaining (36)
radial-velocity members have velocity standard deviations in the
``grey-zone'' between $0.5~km~s^{-1}$ and $2.5~km~s^{-1}$. Orbital parameters
have been determined for 5 of the close binaries. These are marked with
arrows in Figure~\ref{pbv_sm_bm}. The binary primary star
with $P_{rot} = 8.0$ days and $(B-V)_{0} = 1.03$ is in a 4.39 day circular
($e = 0.063 \pm 0.033$) orbit and thus rotates at a sub-synchronous velocity.
The 8 day rotation period is confirmed by \citet{jbm+10}, and in neither
study does a 4.39 day rotation period lead to a well-phased light curve.
This binary and its tidal evolution is discussed in \citet{mms06}. The
binary primary star with $P_{rot} = 4.3$ days and $(B-V)_{0} = 0.69$ is
in a 4.30 day circular ($e = 0.097 \pm 0.006$) orbit, and has synchronized
its rotation to its orbital velocity. The three remaining binaries with
orbits have periods of 280, 715, and 1075 days. We refer to \citet[][Section 5]
{mms06} for a discussion of why the detected photometric variability
represent rotation of the binary primary star. Further study of the
distribution of single and binary primary stars in the color-period
diagram will be presented in a forthcoming paper.

\placefigure{pbv}

\subsection{Comparison with previous work}

Two recent papers present rotation periods for late-type stars in M34
based on time-series photometry. \citet{iah+06} determined rotation periods
for 105 stars and \cite{jbm+10} for 55 stars. Only 58 of the \citeauthor
{iah+06} rotators fall within the color-range of rotators in \citeauthor
{jbm+10} and the present study ($\sim 0.4 \la (B-V)_{0} \la 1.5$). Neither
the \citeauthor{iah+06} nor the \citeauthor{jbm+10} studies have membership
information beyond the results from \citet{jp96}.

We compare here the rotation periods for stars in common between the two
studies and our own. This comparison serves as a test of our own results
as well as to illustrate how {\bf 1)} the time-span and cadence of the
photometric time-series affects the ability to correctly determine the
rotation periods for, in particular, more slowly rotating stars, and
{\bf 2)} erroneous period determinations combined with contamination from
field stars can conspire to obscure the dependencies between stellar
rotation and color (mass).

\begin{figure}[ht!]
\epsscale{1.0}
\plottwo{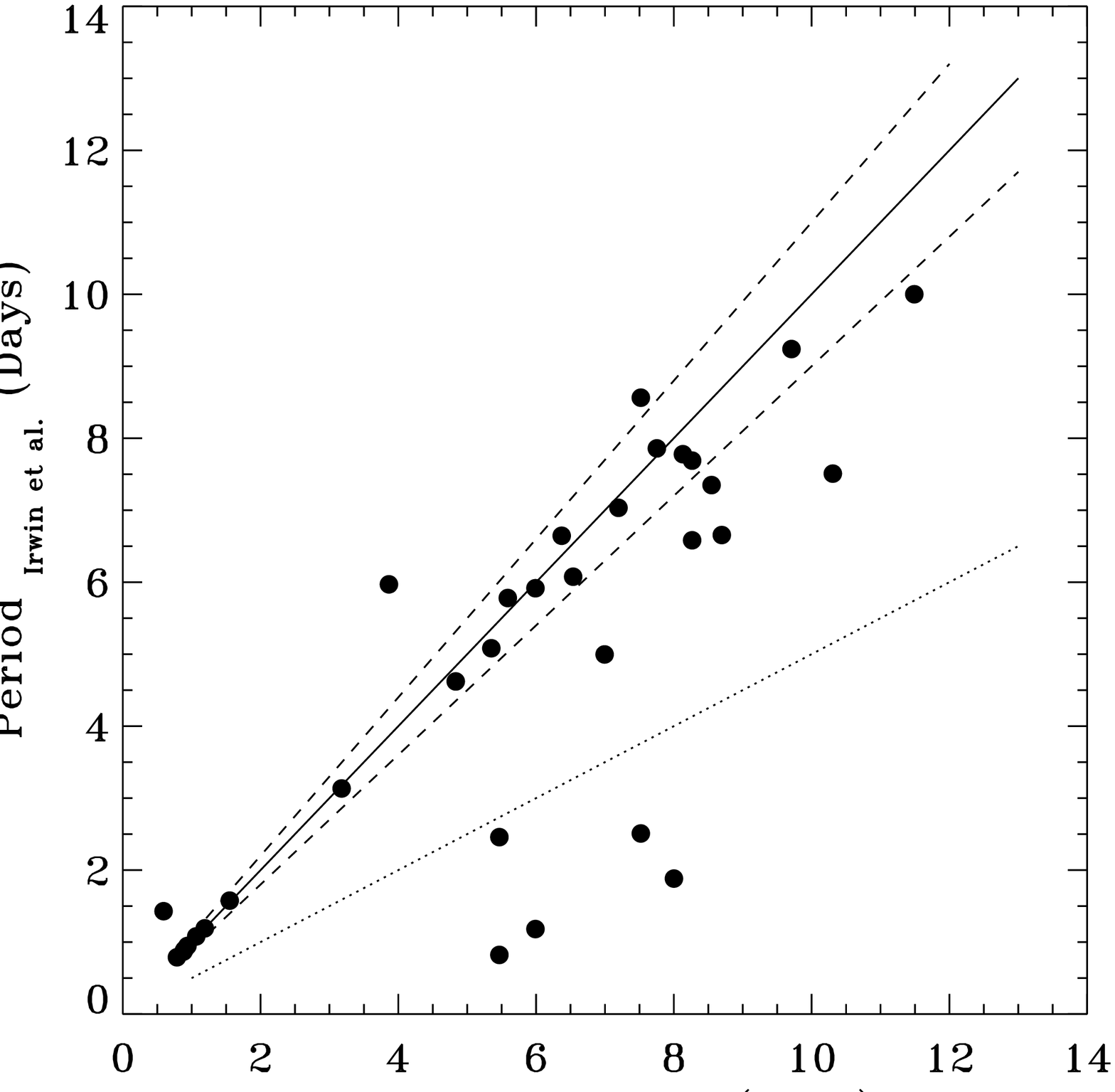}{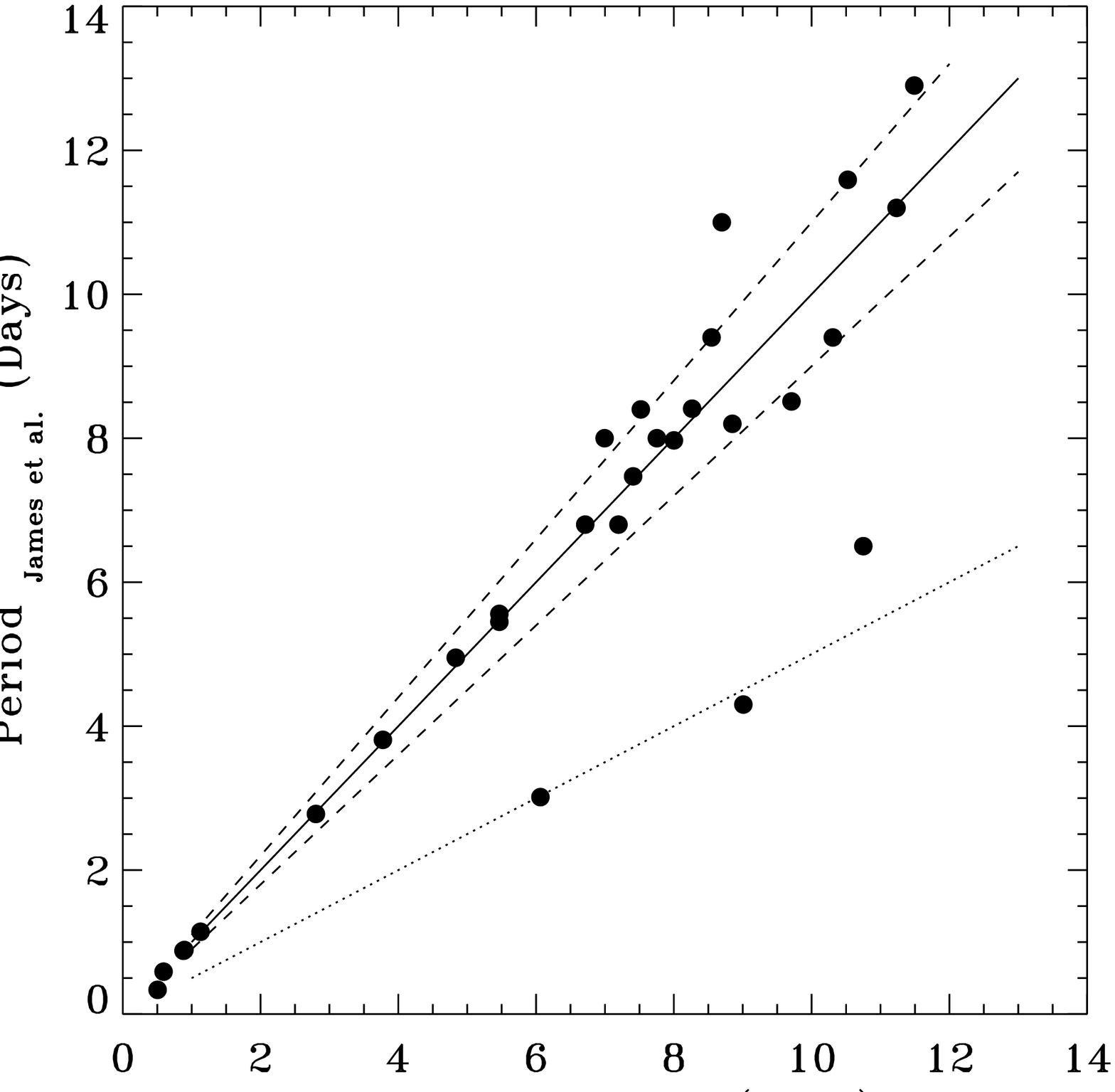}
\caption{A comparison of rotation periods for 33 stars in common
between the present study and \citet{iah+06} ({\it left}) and for 29
stars in common between the present study and \cite{jbm+10} ({\it right}).
The solid and dashed lines mark period equality and 10\% deviations.
The dotted lines mark phase-halving.
\label{p_compare}}
\end{figure}

Figure~\ref{p_compare} shows a direct comparison of rotation periods
for 33 stars ({\it left}) and 29 stars ({\it right}) in common between
the present study and \citet{iah+06} and \citet{jbm+10}, respectively.
Allowing for a 10\% deviation to accomodate period uncertainties, we
find that for periods beyond $\sim$5 days, half of the periods published
by \citeauthor{iah+06} are significantly shorter than ours. This systematic
trend toward shorter periods (bias against longer periods) in the
\citeauthor{iah+06} study is almost certainly due to the nature of their
observations of M34 - half nights over a relatively short 10-night
campaign. We note that the overall photometric precision of the two
studies is similar. The comparison thus demonstrates the importance
of the long time baseline and the frequency of observations to measure the
true periods for the more slowly rotating cluster members.

With a 17-night baseline and observations acquired throughout the entire
nights, the rotation periods determined by \citet{jbm+10} are in better
agreement with ours. Four periods of 6 days and longer deviate significantly,
with 3 of the \citeauthor{jbm+10} periods lying on or close to the
phase-halving line. We too have 3-4 gap stars (see Figure~\ref{pbv}) with
periods near half the period of corresponding stars on the $I$ sequence.
In all three studies, if such periods are indeed half the true periods,
it is likely because of spots at two opposite active longitudes at the time
of the observations that could not be resolved with the sampling and 
photometric precision in the respective datasets.

\begin{figure}[ht!]
\epsscale{1.0}
\plotone{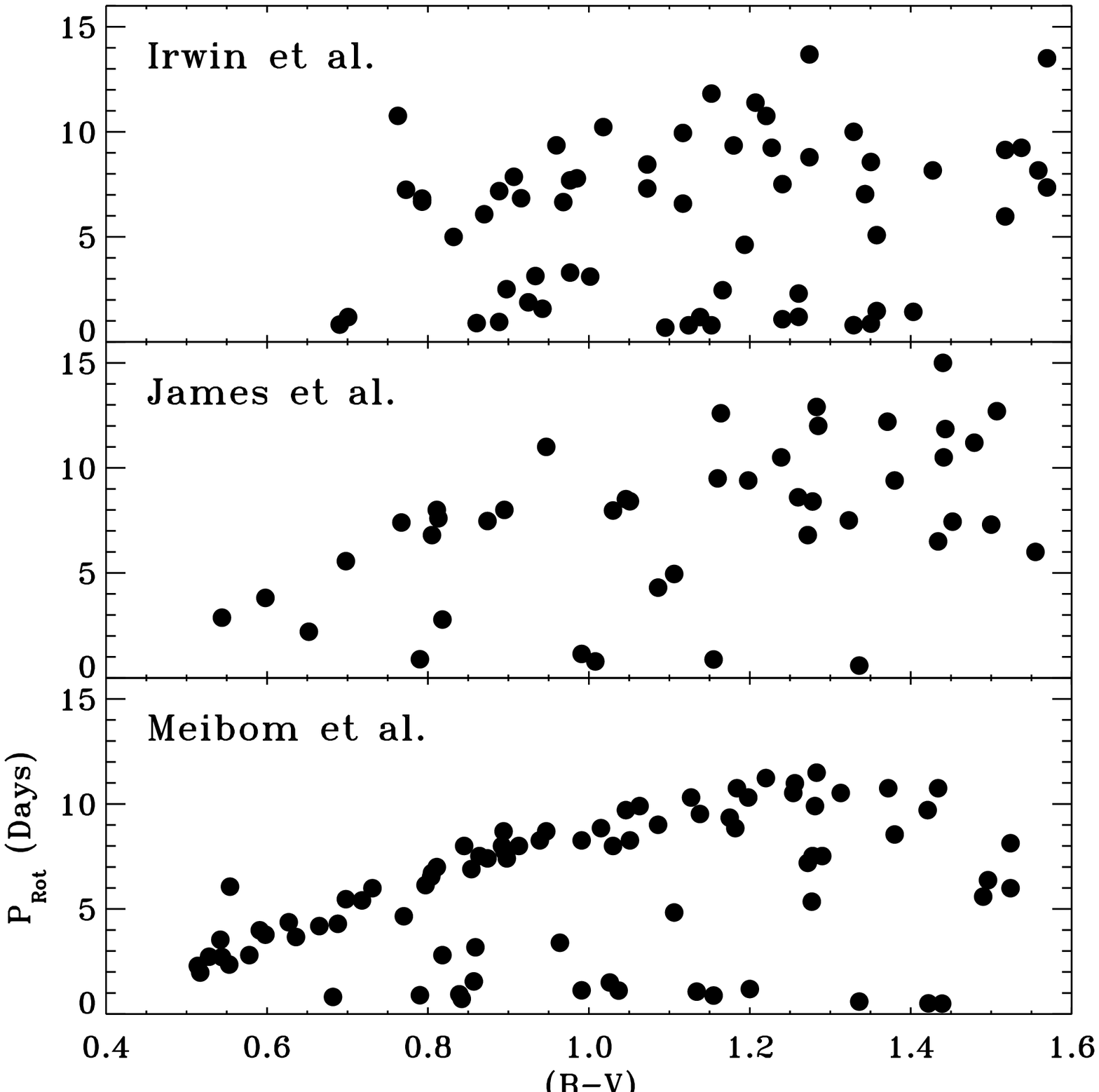}
\caption{The color-period diagrams based on the results from
\citet[][top]{iah+06}, \citet[][center]{jbm+10}, and from the present
study (bottom).
\label{cp_compare}}
\end{figure}

While seemingly innocuous in Figure~\ref{p_compare}, erroneous period
detections combined with small stellar samples and contamination from
field stars can act to obscure the dependencies of stellar rotation
on mass and age revealed in color-period diagrams for open clusters.
This is well illustrated by a comparison of the color-period diagrams
with the results of the three studies of M34. We show in
Figure~\ref{cp_compare} the color-period diagrams based on the results
from \citet[][top]{iah+06}, \citet[][center]{jbm+10}, and from the
present study (bottom). We have used the color-color transformation
by \citet{cca+93} and $E_{V-I} = 1.25 \times E_{B-V}$ to convert the
$V-I$ colors in \citet{iah+06} to $(B-V)_{0}$. We note that the areal
coverage of the \citeauthor{iah+06} and \citeauthor{jbm+10} surveys
are $\sim$75\% and $\sim$50\% of ours. However, the comparison is
qualitative and the number of erroneous period determinations and the
contamination by field stars will increase in proportion to the number
of stars observed (area). We note that the differences between the
current and prior studies are not due to just deviating periods among
cluster members or just contamination by non-members, but rather a
combination of both.

While all three color-period diagrams in Figure~\ref{cp_compare}
capture the large dispersion in stellar rotation among late-type
stars in M34, the striking morphology visible in our new results is
largely hidden in the top and center panels. Previous authors
\citep[e.g.][]{barnes03a,mms09,hgp+09,iab+09,jbm+10} have noted
persistent features in the period-color diagram. The diagonal
sequence of stars on which rotation periods increase with color,
the sequence of rapidly rotating stars, and the region in between,
was dubbed the $I$ sequence, the $C$ sequence, and the {\it gap},
respectively, by \citet[][hereinafter B03]{barnes03a}. We demonstrate
in the sections to follow how the rotational sequences in the
color-period diagram for clusters of different ages enable detailed
study of the angular momentum evolution of cool stars.

\begin{figure}[ht!]
\epsscale{1.0}
\plotone{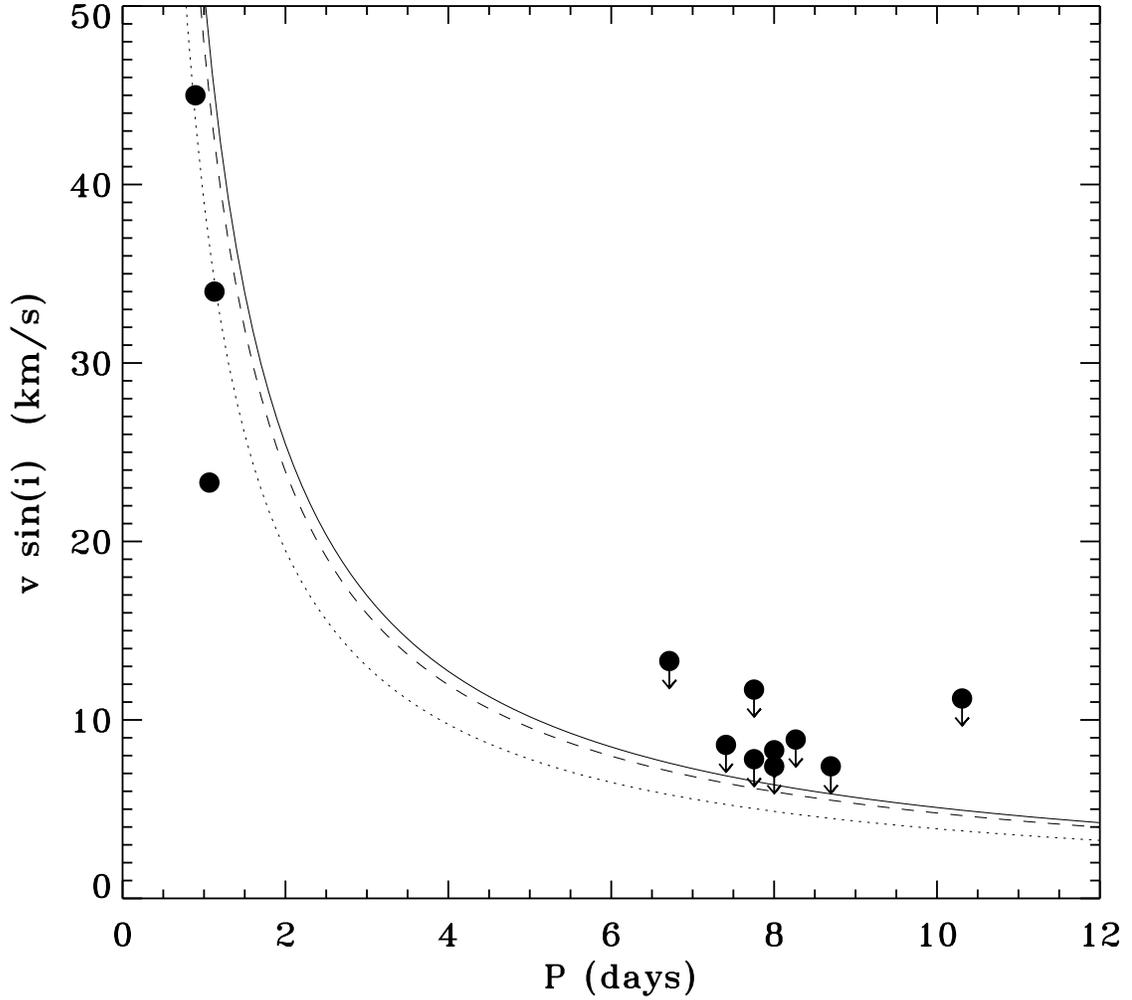}
\caption{Projected rotation velocities \citep{sjf01} plotted
against the measured rotation period for 12 stars in M34. All stars have
radial-velocity cluster membership probabilities larger than 75\% and
none of the 12 stars are spectroscopic binaries. For comparison, the solid,
dashed, and dotted curves indicate the relation between rotation period
and the projected rotational velocity for a solar-like star with a $90\degr$,
$70\degr$, and $50\degr$ inclination of the rotational axis, respectively.
The rotation periods and the projected rotation velocities are consistent
for all 12 stars.
\label{vsini}}
\end{figure}

Finally, Figure~\ref{vsini} shows the $v \sin i$ measurements from
\citet{sjf01} against our rotation periods for 12 members of M34. The solid,
dashed, and dotted curves show the relation between rotation period and the
$v \sin i$ for a solar-like star with a $90\degr$,
$70\degr$, and $50\degr$ inclination of the rotational axis, respectively.
The 3 stars with rotation periods of $\la 1$ day have correspondingly high
$v \sin i$ values.

\section{ANALYSIS AND DISCUSSION}
\label{analysis}

The color-period diagrams for young open clusters like M34 have emerged
as important tools for describing the rotational properties of low-mass
stars after the PMS phase \citep[e.g.][]{iab+09,mms09,hgp+09,hbk+10,jbm+10}.
The mass- and time-dependent changes of the sequences in the color-period
diagram therefore makes it an important testing ground for models
of stellar angular momentum evolution - much like the Hertzsprung-Russell
diagram for models of stellar evolution. Here we compare the M34 color-period
diagram to those of younger and older clusters with the goal of estimating
the timescale and rate for spin-down of the stellar surface and its dependence
on stellar mass.


\subsection{Timescales for transition from fast to slow rotation}
\label{cet}

\begin{figure}[ht!]
\epsscale{0.7}
\plotone{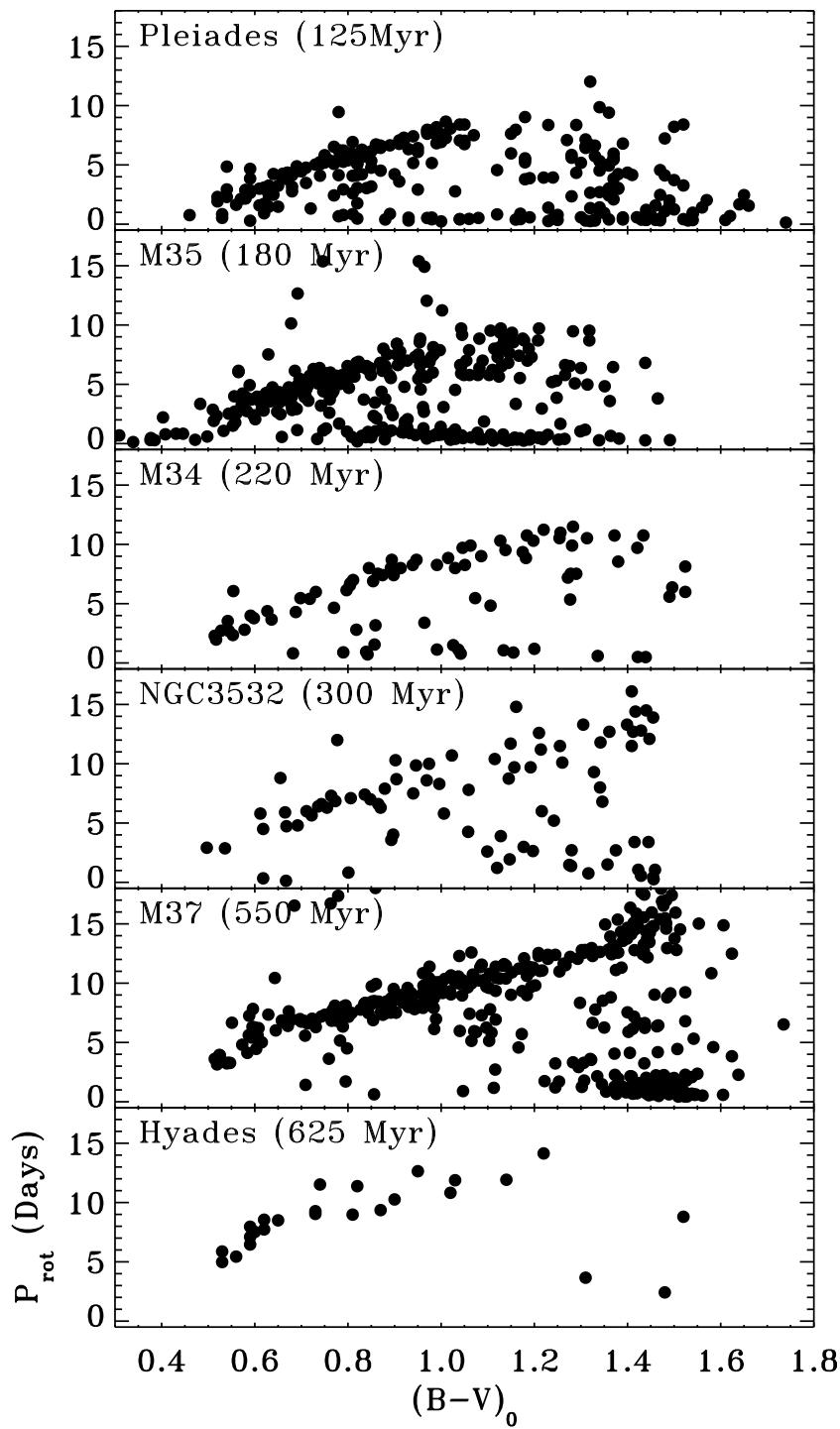}
\caption{The color-period diagrams for the Pleiades \citep[][only stars with
measured B-V colors are shown]{hbk+10}, M35 \citep{mms09}, M34, NGC3532 (B03),
M37 \citep{hgp+09} and the Hyades \citep{rtl+87}. The cluster names and ages
are given in upper left of each panel.
\label{cps}}
\end{figure}

In Figure~\ref{cps} we show the color-period diagram for M34 with the
color-period diagrams for the Pleiades (125\,Myr), M35 (180\,Myr),
NGC3532 (300\,Myr), M37 (550\,Myr), and the Hyades (625\,Myr). The
diagrams are shown in chronological order with the youngest cluster
in the top panel. All clusters display distinct $I$ and $C$ sequences
and sample well the age range between the ZAMS and the age of the Hyades.
We can use this line-up of color-period diagrams to estimate the timescale
for the rotational evolution of stars off the $C$ sequence and onto the $I$
sequence for stars of different masses.

Already by the age of the Pleiades and M35 the number of rapidly rotating
G dwarfs on the $C$ sequence is small, whereas the G dwarfs $I$ sequences
are rich and well defined. Observations in the even younger clusters IC2391
\citep[$\sim$30\,Myr][]{ps96}, IC2602 \citep[$\sim$30\,Myr][]{bsp+99}, and
Alpha Per \citep[$\sim$50\,Myr][]{psm+93} find $C$ sequences well
populated by G dwarfs \citep[see][]{barnes03a}. Together,
these results suggest that the characteristic timescale for G dwarfs to
evolve off the $C$ sequence and onto the $I$ sequence is less than
$\sim$150 Myr. By the $\sim$300\,Myr age of NGC3532, the rich populations
of early- and mid-K dwarfs found on the M35 $C$ sequence have evolved onto
the $I$ sequence, while the NGC3532 $C$ sequence and gap are populated by
late K dwarfs. This change suggests that early to mid K dwarfs evolve onto
the $I$ sequence on a timescale between 150 and 300 Myr, or approximately
twice the time required for G dwarfs. Finally, by the age of M37 and the
Hyades, mostly early M dwarfs are found on the $C$ sequence or in the gap,
suggesting that late-K dwarfs evolve off the $C$ sequence and onto the $I$
sequence within $\sim$600\,Myr, or approximately twice the time required
for the early to mid K dwarfs.

\placefigure{cps}

\clearpage

The low density of stars in the gap region for all of these young clusters
indicates that stars spend less time in this part of the diagram than on
the $C$ and $I$ sequences (a feature of the color-period diagram that is
similar to the Hertzsprung-gap in the color-magnitude diagram). The narrow
$C$ and $I$ sequences in both the Pleiades and M35 also suggest that the
rapidly rotating $C$ stars lose little angular momentum until they quickly
evolve through the gap and onto the $I$ sequence. Accordingly, their
surface spin-down rate ($dP/dt$) must reach a maximum value in the gap.
We can derive a rough estimate of $(dP/dt)_{gap}$ for K dwarfs from the
$I$ and $C$ sequences for the Pleiades. To that end we make the assumption
that the Pleiades $C$ sequence represent the rotation periods for K dwarfs
at the ZAMS and that the periods of the Pleiades K dwarfs on the $I$
sequence are representative of post-gap rotation. Accordingly, the
early K $I$ sequence dwarfs with periods of $\sim$6\,days and the late K
$I$ sequence dwarfs with periods of $\sim$8\,days have spun down from the
$\sim$0.5\,day periods over $\sim$100\,Myr. This corresponds to
$(dP/dt)_{gap}$ of $\sim$0.06\,days/Myr and $\sim$0.08\,days/Myr
for early and late K dwarfs, respectively. These are crude estimates,
as described, and the spin-down rate will not be constant as a star
transition through the gap. The maximum $(dP/dt)_{gap}$ will likely
be higher than the derived values.

The timescales for the transition from fast ($C$ sequence) to slow ($I$
sequence) surface rotation as well as the lower limit on the maximum
spin-down rate though the gap, offer constraints on the rates of internal
and external angular momentum transport and on the evolution of stellar
dynamos in late-type stars of different masses.


\subsection{Testing the Skumanich $\sqrt t$ Spin-Down Rate Between M34 and the Hyades}
\label{testing_skumanich}

\begin{figure}[ht!]
\epsscale{1.0}
\plottwo{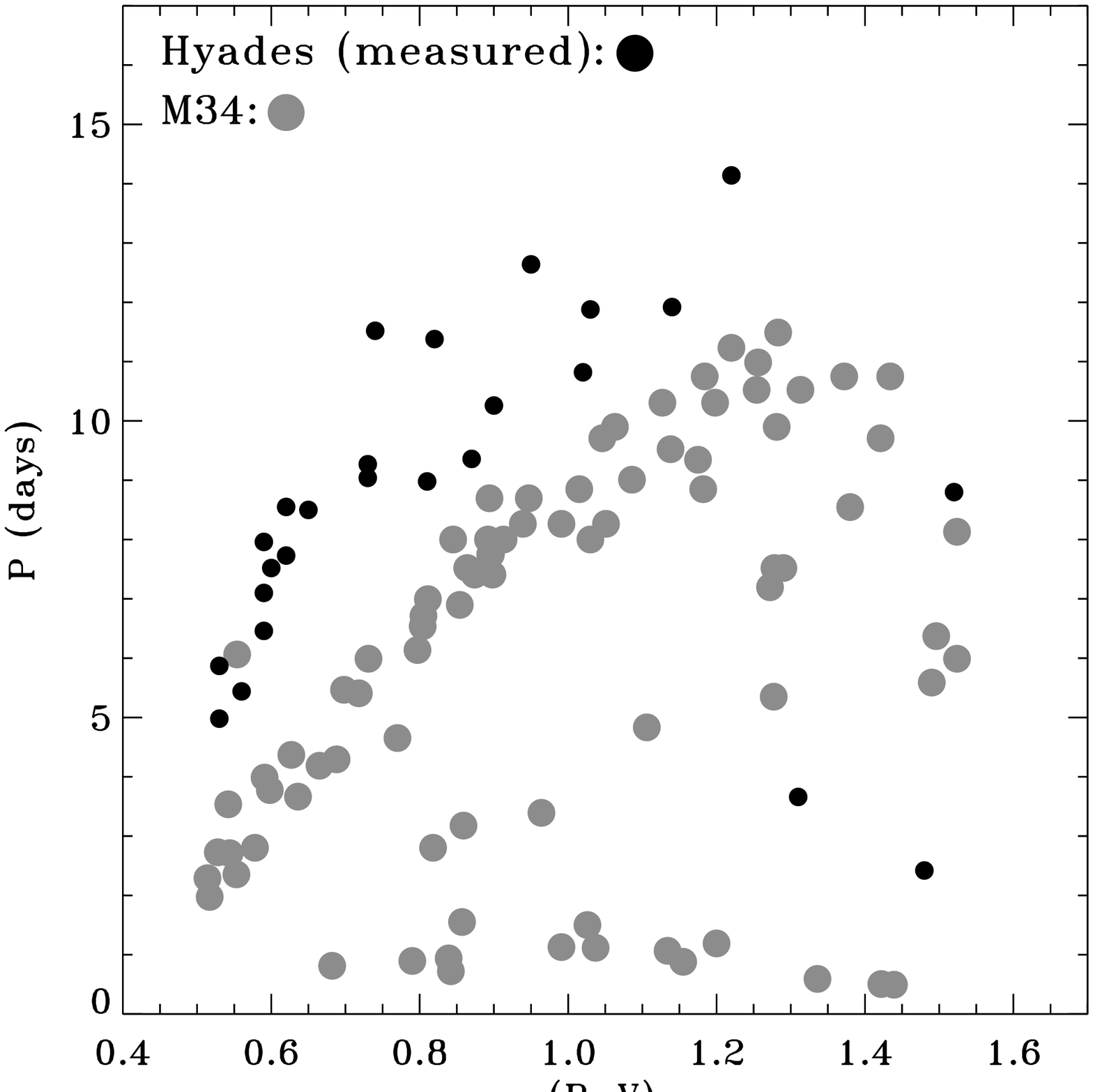}{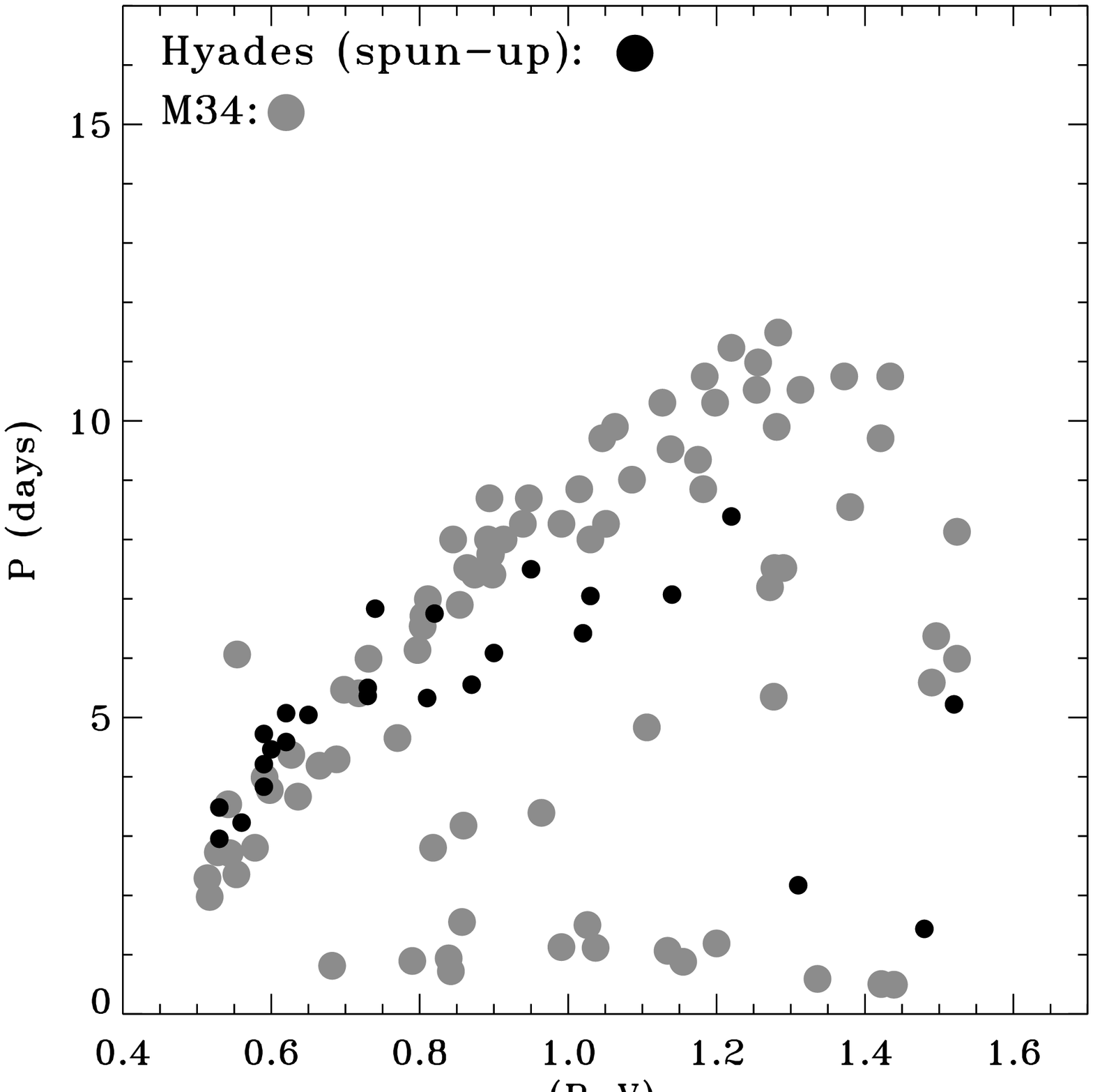}
\caption{{\it Left:} Color-period diagram with results for M34 (220\,Myr)
and the Hyades (625\,Myr).
{\it Right:} The rotation periods for the 25 Hyades stars are spun-up by
a factor $\sqrt{625/220}$ to test the $\sqrt{t}$ Skumanich spin-down law.
\label{m34_hya}}
\end{figure}

We show in Figure~\ref{m34_hya} (left) the color-period diagrams for M34
and the Hyades. The rotational evolution over the $\sim$400\,Myr age-gap
is clearly visible with the Hyades $I$ sequence G and K dwarfs offset to
longer periods. Also, the 3 late K/early M Hyades $C$ sequence dwarfs
appears to have evolved into the gap.
Once stars have converged in their rotational evolution onto the $I$
sequence it is assumed that they steadily lose angular momentum through
a solar-type wind and spin-down following the Skumanich $\sqrt{t}$
spin-down law. Indeed, \citet{skumanich72} used $v \sin i$ measurements
for early-G $I$ sequence stars in the Pleiades and Hyades. With a
well-defined $I$ sequence extending to late-K dwarfs in M34 we can
directly test whether the evolution of the surface rotation period for
K dwarfs follow the $\sqrt{t}$ time-dependence between the ages of M34
and the Hyades. In the right panel of Figure~\ref{m34_hya} we have
artificially spun up the Hyades stars by a factor of $\sqrt{625/220}$
in accordance with the Skumanich spin-down law. The spun-up Hyades $I$
sequence G dwarfs coincide with the corresponding M34 stars, while
the Hyades $I$ sequence K-dwarfs have rotation periods systematically
shorter than the M34 K dwarfs. We conclude that while the G dwarfs
follow the Skumanich spin-down law, the K-dwarfs spin down slower
than the $\sqrt{t}$ rate from 220\,Myr to 625\,Myr. A similar result
was found from our comparison of M35 (180\,Myr) with the Hyades (Paper I).

\placefigure{m34_hya}

\subsection{Gyrochronology and the gyro-age of M34}
\label{gyro}

The $I$ sequence in the color-period diagram, onto which essentially all
FGK dwarfs evolve within $\sim$600\,Myr (see Figure~\ref{cps}), provide
the basis for establishing empirically the relationship between stellar
rotation and stellar color (mass) at a given age. Accordingly, color-period
diagrams for clusters of different ages can be used to establish the
relationships between stellar rotation and stellar age for stars of
different masses.

B03 introduced, and \citet[][hereinafter B07]{barnes07} modified, a heuristic
functional form ($P(t,B-V) = g(t) \times f(B-V)$) to represent the 
dependency of the $I$ sequence. $P(t,B-V)$ define one-parameter families,
with that parameter being the age of the stellar population, and the
resulting curves in the color-period plane represent rotational isochrones.
B07 proposed 

\begin{equation}
P(t,B-V) = g(t) \times f(B-V) = t^{n} \times \bigl(a((B-V)-b)^{c}\bigr)
\end{equation}

\noindent with values for the $a$, $b$, $c$, and $n$ coefficients determined
using the $I$ sequences from multiple young open clusters. A value of 0.52
was determined for the $n$ index in $g(t)$ by demanding solar rotation at
solar age. We note that \citet{mh08} determined different coefficients in
the gyrochronology relation than B07.

In Paper I we used the tight M35 $I$ sequence to redetermine the $a$, $b$,
and $c$ coefficients in $f(B-V)$. Here, we wish to use $f(B-V)$ determined
from the M35 $I$ sequence to determine a new gyrochronology age (gyro-age)
for M34. Furthermore, to illustrate the effect of an uncertainty in the
age of M35 on the gyro-age for M34, we determine $f(B-V)$ from M35 using
three different ages for that cluster that span the range set by modern
photometric studies: 160\,Myr \citep{ssd+00}, 180\,Myr \citep{kfr+03},
and 200\,Myr \citep{sb99}. Table~\ref{coeffs} lists the $a$, $b$, and $c$
coefficients for $f(B-V)$ resulting from non-linear least squares fits of
$P(t,B-V)$ to the M35 $I$ sequence with $t = 160\,Myr$, $t = 180\,Myr$,
and $t = 200\,Myr$, respectively. Note that only the $a$ coefficient is
affected significantly by a change in age.

\begin{deluxetable}{ccccc}
\tabletypesize{\normalsize}
\tablecaption{Coefficients (with $1 \sigma$ uncertainties) for the B07
$I$ sequence rotational isochrones ($P(t,B-V) = t^{0.52} \times
a((B-V)-b)^{c}$) based on least squares fits to $I$ sequence stars
in M35 (paper I) using $t = 160\,Myr$, $t = 180\,Myr$, and $t = 200\,Myr$.
Fourth column lists the coefficients determined from a least squares fit
of $P(t,B-V)$ to the M34 $I$ sequence using $t = 220\,Myr$.
\label{coeffs}}
\tablewidth{0pt}
\tablehead{
\colhead{Coefficient} & \colhead{M35} & \colhead{M35} & \colhead{M35} & \colhead{M34}\\
\colhead{} & \colhead{$t = 160\,Myr$} & \colhead{$t = 180\,Myr$} & \colhead{$t = 200\,Myr$} & \colhead{$t = 220\,Myr$}\\
}
\startdata
a & $0.744\pm 0.013$ & $0.700\pm 0.013$ & $0.663\pm 0.012$ & $0.730\pm 0.013$\\
b & $0.472\pm 0.027$ & $0.472\pm 0.027$ & $0.472\pm 0.027$ & $0.481\pm 0.017$\\
c & $0.553\pm 0.052$ & $0.553\pm 0.052$ & $0.553\pm 0.052$ & $0.532\pm 0.041$\\
\enddata
\end{deluxetable}

\begin{figure}[ht!]
\epsscale{1.0}
\plottwo{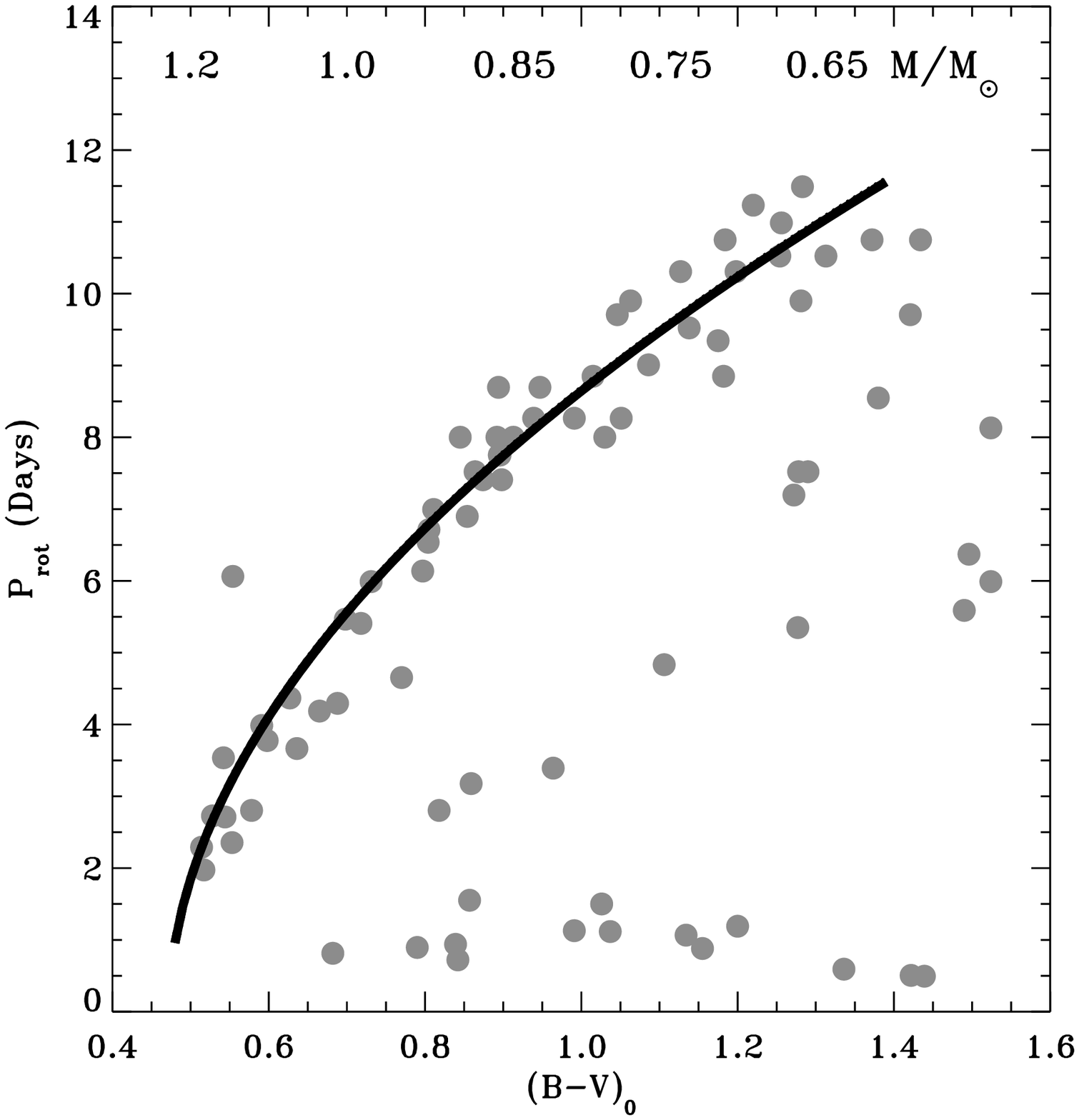}{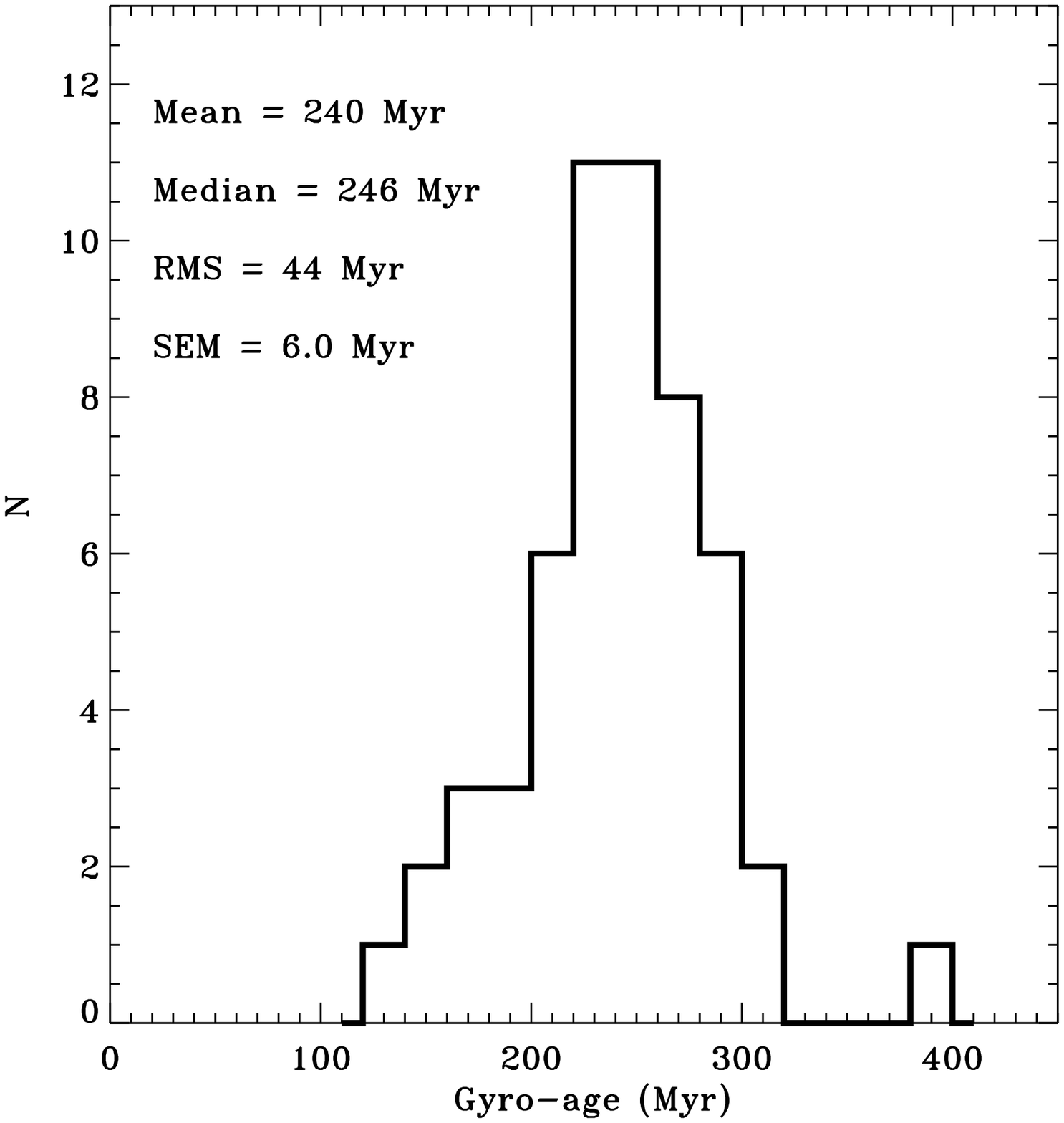}
\caption{{\bf Left:} The location on the M34 $I$ sequence of the best
fit rotational isochrones. All three isochrones fall within the black
solid curve. 
{\bf Right:} The distribution of gyro-ages for all M34 $I$ sequence
stars calculated using $P(t,B-V)$ with coefficients determined
from the M35 $I$ sequence with an assumed age of 180\,Myr.
The distribution mean, median, standard deviation, and standard error
on the mean are given in the upper left.
\label{gyroages}}
\end{figure}

Using the three sets of coefficients (and $n = 0.52$) and assuming no
prior knowledge of the age of M34 (letting $t$ be a free parameter), we
perform a non-linear least squares fit of $P(t,B-V)$ to the M34 $I$
sequence stars. The best fits for the three different isochrones all fall
within the black solid curve shown in the left panel of Figure~\ref{gyroages}.
The corresponding gyro-ages are $213 \pm 4\,Myr$, $240 \pm 4\,Myr$, and
$266 \pm 5\,Myr$, respectively, for assumed M35 ages of 160\,Myr, 180\,Myr,
and 200\,Myr. The range in gyro-ages for M34 directly reflects the assumed
uncertainty on the age of M35 through the coefficients in $f(B-V)$. It
corresponds to $\pm10\%$ deviations from the mean gyro-age of 240\,Myr
that results from using an age of 180\,Myr for M35 - the age adopted for
this paper. For comparison, B07 estimate the representative errors associated
with the gyrochronology technique to $\sim$15\% based on a more
sophisticated analysis of errors in the measured stellar rotation periods,
colors, and cluster ages. The range in M34 gyro-ages overlaps with the
range of stellar evolution ages for the cluster and include the mean
stellar evolution age of 220\,Myr adopted for this paper. We conclude
that the gyro-ages derived for M34 are consistent with the stellar
evolution age for the cluster, and that the measured effect of age
uncertainties for M35 on the M34 gyro-age is consistent with the
error analysis by B07. We adopt the gyro-age of 240\,Myr for M34
corresponding to an age of 180\,Myr for M35.

We emphasize the small formal ($1\sigma$) error on the gyrochronology
age of M34. Assuming that all $I$ sequence stars are truly coeval,
the $\sim$5\,Myr standard error gives a formal uncertainty on the
gyro-age for M34 of only 2\%. This small formal uncertainty, which
does not include possible systematic errors in gyrochronology, reflect
the well-defined $I$ sequence in M34. We show in the right panel of
Figure~\ref{gyroages}, the distribution of gyro-ages for all M34
$I$ sequence stars calculated using $P(t,B-V)$ with coefficients
determined from M35 with an assumed age of 180\,Myr. The mean and median
ages of the distribution are 240\,Myr and 246\,Myr, respectively. The
most deviant gyro-ages in the distribution ($< 180\,Myr$ and $> 320\,Myr$)
are for late F and early G type stars near the blue and steepest part
of the $I$ sequence, where the deviation from the rotational isochrone
is a larger fraction of the stellar rotation period.

\placefigure{gyroages}

\begin{figure}[ht!]
\epsscale{1.0}
\plotone{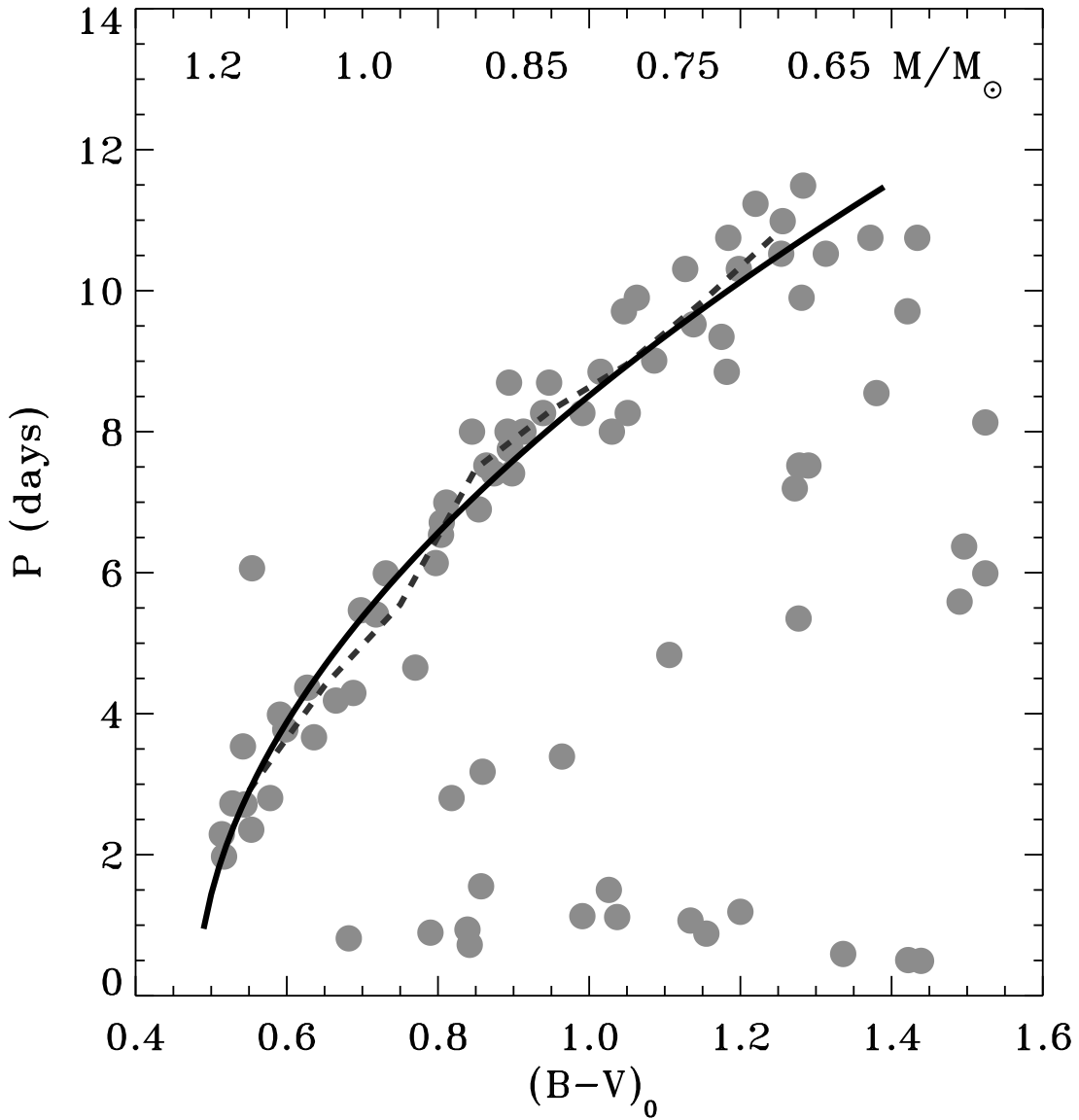}
\caption{The least squares fit (solid curves) of 220\,Myr B07 $I$ sequence
isochrones (eq. [1]) to the M34 $I$ sequence. The corresponding new values
for the isochrone coefficients $a$, $b$, and $c$, are listed in
Table~\ref{coeffs}. The moving average of the rotation periods for the
$I$ sequence stars is also shown as a dashed curve for comparison.
\label{newiso}}
\end{figure}

As demonstrated, the method of gyrochronology relies on fitting
the $I$ sequence with a rotational isochrone. The isochrone's functional
dependence between stellar color and rotation period will thus directly affect
the derived gyro-age. To constrain the color-period relation further we
redetermine the $a$, $b$, and $c$ coefficients of Equation [1] from a
non-linear least squares fit to the M34 $I$ sequence stars with a fixed cluster
age ($t$) of 220\,Myr. The coefficients with $1\sigma$ uncertainties are
listed in Table~\ref{coeffs} and the corresponding rotational isochrone
is shown in Figure~\ref{newiso}. To illustrate how closely the isochrone
trace the $I$ sequence, we plot a dashed curve representing the moving
average of the rotation periods along the $I$ sequence.

\placefigure{newiso}

The translational term $b$ in Equation 1, fixed to 0.4 by B07 to allow
for more blue stars to be fitted, corresponds to the $B-V$ color for
F-type stars at the transition from a radiative to a convective envelope.
This transition was noted from early observations of stellar rotation
(the break in the Kraft curve, \citet{kraft67}) and is associated with
the onset of effective magnetic wind breaking \citep[e.g.][]{schatzman62}.
The fitted values of $b$ for M35 and M34 suggest that blue (high-mass)
end of the $I$ sequence begins at the break in the Kraft curve and set
the $B-V$ color of the break to 0.472 and 0.481, respectively.


\subsection{Long-term stability of stellar spots and spot-groups}

The majority of the 120 phased light curves shown in Appendix A
(Figure~\ref{plc}), for both
cluster and field stars, display a notable stability in the phase,
shape, and amplitude of photometric variability over the $\sim$5 month
duration of our survey. We discuss here two possible explanations for
this stability: 1) Spots on the observed stars are significantly
longer lived than sunspots, or, 2) Spots on the observed stars tend to
emerge non-uniformly at preferred long-lived active longitudes.

Let us first examine the idea of long-lived starspots. On the Sun,
larger sunspots live longer, with lifetimes increasing linearly with
area. To first order, however, the solar log-normal spot size
distribution is strongly dominated by small spots and largely unchanged
(except for an overall scaling factor) over the solar cycle \citep{bgl+88}.
Even permitting a small fractional increase in larger spots at cycle
maximum, allowed by the solar data, \citet{su04} found that when extrapolating
to very active stars, small spots below the resolution limit still dominate.
Furthermore, the enhanced differential rotation expected in the more rapidly
rotating stars \citep[$\Delta P_{\rm rot}/\langle P_{\rm rot} \rangle
\propto P_{\rm rot}^{0.3}$;][]{dsb96} would tend to {\it decrease}
spot lifetimes due to increased shear. Thus we do not expect that an
increase in numbers of spots will significantly change the average
spot lifetime on stars.

We also note that there is no direct evidence that the solar log-normal
spot size distribution holds for other stars. However, there {\it is}
evidence that small spots dominate stellar surfaces. Measurement
of TiO band depths
\citep[e.g.][]{osn96} imply that a large number of unresolved spots are
present on active stars, and high precision photometric measurements during
stellar \citep{jbc+06} and exoplanet eclipses \citep{pgm+07,lpl+09,sla+10}
indicate spots and spot groups similar in size to solar. Finally, there
is also direct evidence that starspot lifetimes are short \citep{mbl+09}
even if the lifetimes of large-scale spot groups can seem long on similar
stars \citep[e.g.][]{hussain02}.

To reconcile these apparently disparate observations, the second
possibility can be invoked: starspots emerge at preferred longitudes.
If the ``spots'' seen by most photometric (and Doppler imaging
observations) are actually mostly spot {\it groups} consisting of
a range of smaller spots emerging at preferred longitudes, the observed
spottedness can be maintained for long periods even if the underlying
individual spots come and go.  The only requirement is that the emergence
rate is sufficiently high to maintain a quasi-steady ``spot'' size,
but this is expected - at least on stars as young and active as those
in M34. The Sun too shows evidence for preferred longitudes \citep{bu03},
but the phenomenon is widespread in active stars \citep{kj07}.  While
seen most often in binaries, preferred longitudes are also observed
in single stars \citep{bj05}.  We therefore suggest that the five month
stability of rotational patterns on most of our objects is due to
enhanced spot generation rates coupled with preferred longitudes
of spot emergence.


\section{SUMMARY AND CONCLUSIONS}
\label{conclusions}

We present the results of an extensive time-series photometric survey
over $\sim$5 months of late-type members in the 220\,Myr open cluster
M34. We obtain a photometric precision of $\sim$0.5\% and light curves
for 5,656 stars with $12 \la V \la 20.8$ over a $40\arcmin \times 40\arcmin$
field on the cluster. We measure surface rotation periods for 120 stars
and determine their cluster membership and binarity from the results of
a 4-year radial-velocity survey in M34 and from published proper-motion
measurements. The result is surface rotation periods for 83 kinematic and/or
photometric cluster members.

The rotation periods of the 83 cluster members span over more than an
order of magnitude from 0.5 day up to 11.5 days. Plotted as a function
of stellar color in the color-period diagram, the 83 periods provide an
exceptionally clear definition of the relationship between stellar
rotation and stellar mass at 220\,Myr. A comparison of our periods with
those of recently published photometric surveys in M34 underscores the
importance of high frequency and long time-baseline observations to
minimize false period determinations, and of cluster membership
information to minimize field star contamination.

The 83 rotation periods in M34 trace two distinct rotational sequences
in the color-period diagram, representing two different states in their
rotational evolution. Color-period diagrams for select younger
and older clusters show similar sequences, and we compare these with M34
to study the mass-dependent timescales and rates of the spin-down
of late-type stars. 
Analysis of the chronologically sorted color-period diagrams supports the
idea that late-type stars evolve from fast to slow rotation on a timescale
that is inversely related to the stellar mass. We estimate the timescales
for the rotational evolution off the $C$
sequence and onto the $I$ sequence for G, early-mid K, and late-K dwarfs,
respectively, to be $\la$150\,Myr, $\sim$150-300\,Myr, and $\sim$300-600\,Myr.
We estimates the spin-down rate ($dP/dt$) through the gap between the
sequences to $\sim$0.06 days/Myr and $\sim$0.08 days/Myr for early
and late K dwarfs, respectively. These timescales and rates offer constraints
on the rates of internal and external angular momentum transport and on the
evolution rates of stellar dynamos in late-type stars of different masses.

A comparison of the M34 color-period diagram with that of the older Hyades
clusters, confirms our finding in Paper I of mass-dependent deviations
from the Skumanich $\sqrt{t}$ spin-down rate on the main sequence. The K
dwarfs spin down slower than the $\sqrt{t}$ rate from M34 (220\,Myr)
to the Hyades (625\,Myr).

We use rotational isochrones constrained by the $I$ sequence in M35
(Paper I) to determine a new gyrochronology age for M34 of 240\,Myr
from a fit to the M34 $I$ sequence. We find that an assumed $\sim$10\%
age-uncertainty for M35 propagates into a $\sim$10\% uncertainty on
the gyro-age for M34 and conclude that the this effect is consistent
with the systematic errors on gyrochronology ages estimated by
\citet{barnes07}. The range of gyro-ages found for M34 includes the
mean stellar evolution age for the cluster of 220\,Myr. We emphasize
the small formal uncertainty of 2\% on the gyro-age of M34. The small
formal error reflects the well-defined $I$ sequence in M34. We use
the narrow $I$ sequence in the M34 color-period diagram to further
adjust the coefficients in the color-period relation for late-type
stars.

Finally, we comment on a high level of stability in the phase, shape,
and amplitude of the photometric variability of the 120 rotators over
the $\sim$5 month duration of our survey. We propose that the observed
stability is due to enhanced spot generation at active stellar longitudes.

We conclude by emphasizing that the color-period diagram for open star
clusters can reveal sequences whose mass- and time-dependent changes
constitute a crucial testing ground for future models of stellar angular
momentum evolution - much like the color-magnitude diagram for models
of stellar evolution.


\acknowledgments

We wish to thank NOAO and the University of Wisconsin - Madison Department
of Astronomy for the time granted on the WIYN 0.9m and 3.5m telescopes. We
express our deep appreciation to the site managers and support staff at both
telescopes for their exceptional and friendly support. We are thankful to
all observers in the WIYN 0.9m consortium who provided us with high-quality
data through the queue-scheduled observing program. We thank Imants Platais
and Constantine Deliyannis for providing astrometric and photometric data to
support in support of this study. This work has been supported by partial
support to S.M. from NASA grant NNX09AH18A (The Kepler Cluster Study),
from partial support to S.M. from the Kepler mission via NASA Cooperative
Agreement NCC2-1390, NSF grants AST-0349075 and AST-0909365 to KGS, and NSF
grants AST-0406615 and AST-0908082 to RDM. Finally we thank the anonymous
referee for valuable comments and suggestions that improved the presentation
and analysis of our results.

\clearpage



\appendix

\section{PHASED LIGHT CURVES}

\begin{figure}
\epsscale{.92}
\plotone{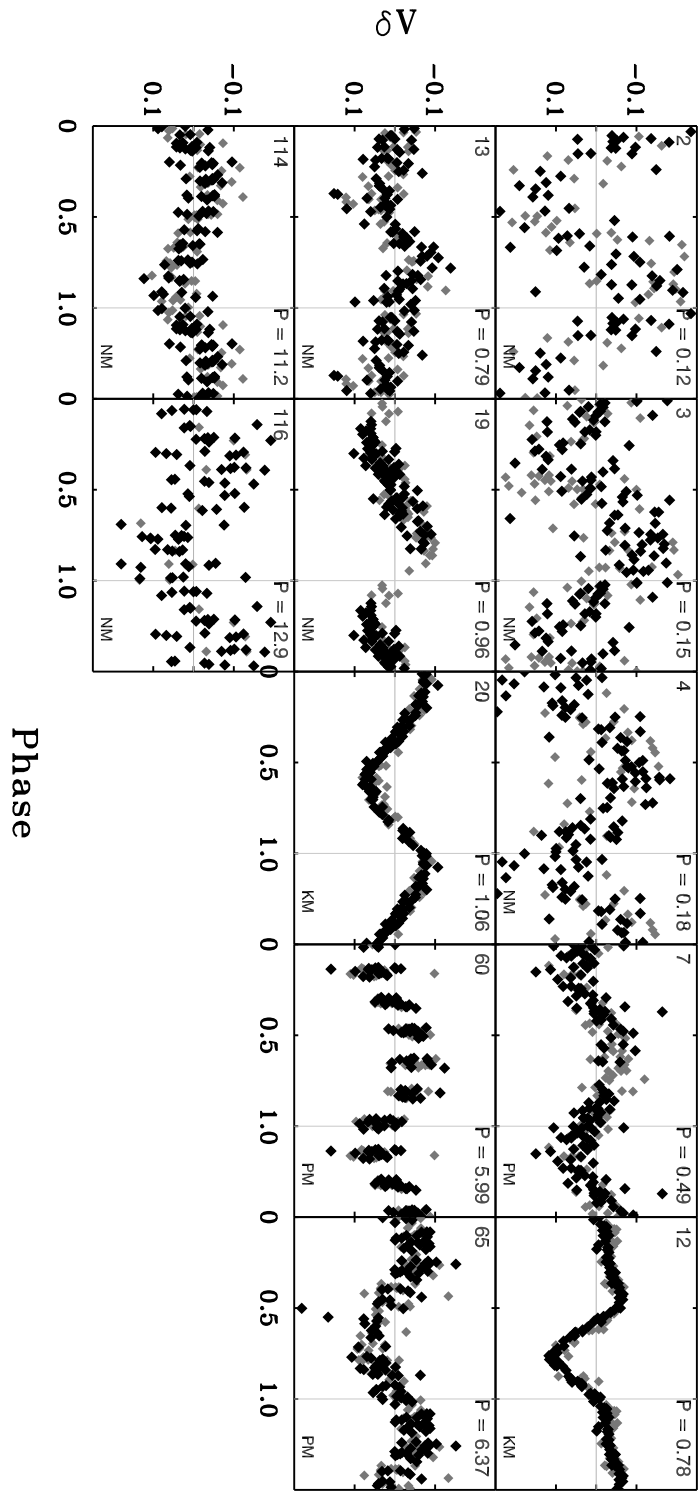}
\caption{Phased light curves for stars with measured rotation periods.
Phased light curves for all 120 stars can be found in the electronic
edition of the Journal.
\label{plc}}
\end{figure}

This appendix presents the light curves for the stars in the field of M34
for which we measured rotation periods. In the printed journal,
Figure~\ref{plc} shows an example of the phased light curve plots.
Phased light curves for all 120 stars can be found in the
electronic edition of the Journal. The light curves have been divided
into 3 groups according to the amplitude of the photometric variation.
For each group the light curves are sorted by the rotation period and
are presented with the same $\delta V$ range on the ordinate. The group
of stars with the largest photometric variability are shown first (in the
electronic edition of the Journal).

For each star we plot the data from the high-frequency survey (December 2002)
as black symbols and data from the low-frequency survey (October 2002 through
March 2003) as grey symbols. A running ID number corresponding to the ID
number in Table~\ref{tab2} in Appendix B is given in the upper left hand
corner in each plot. The period to which the data are phased (the rotation
period listed in Table~\ref{tab2} as $P_{rot}$) is given in the upper right
corner. The letter code in the lower right corner informs about the stars
membership status. The codes have the following meaning: Photometric Member
(PM; described in Section~\ref{phot}), Kinematic (and Photometric) Member
(KM), Non-Member (NM). In each plot a horizontal grey line mark $\delta V =
0.0$ and a vertical grey line marks a phase of 1.0.

\notetoeditor{For the printed edition of the journal, the second panel
(f14b.eps) of Figure 14 should be used and could be shrunk slightly and
rotated 90 degrees to fit below the text above}

\placefigure{plc}

\clearpage
\pagestyle{empty}

\clearpage
{\plotone{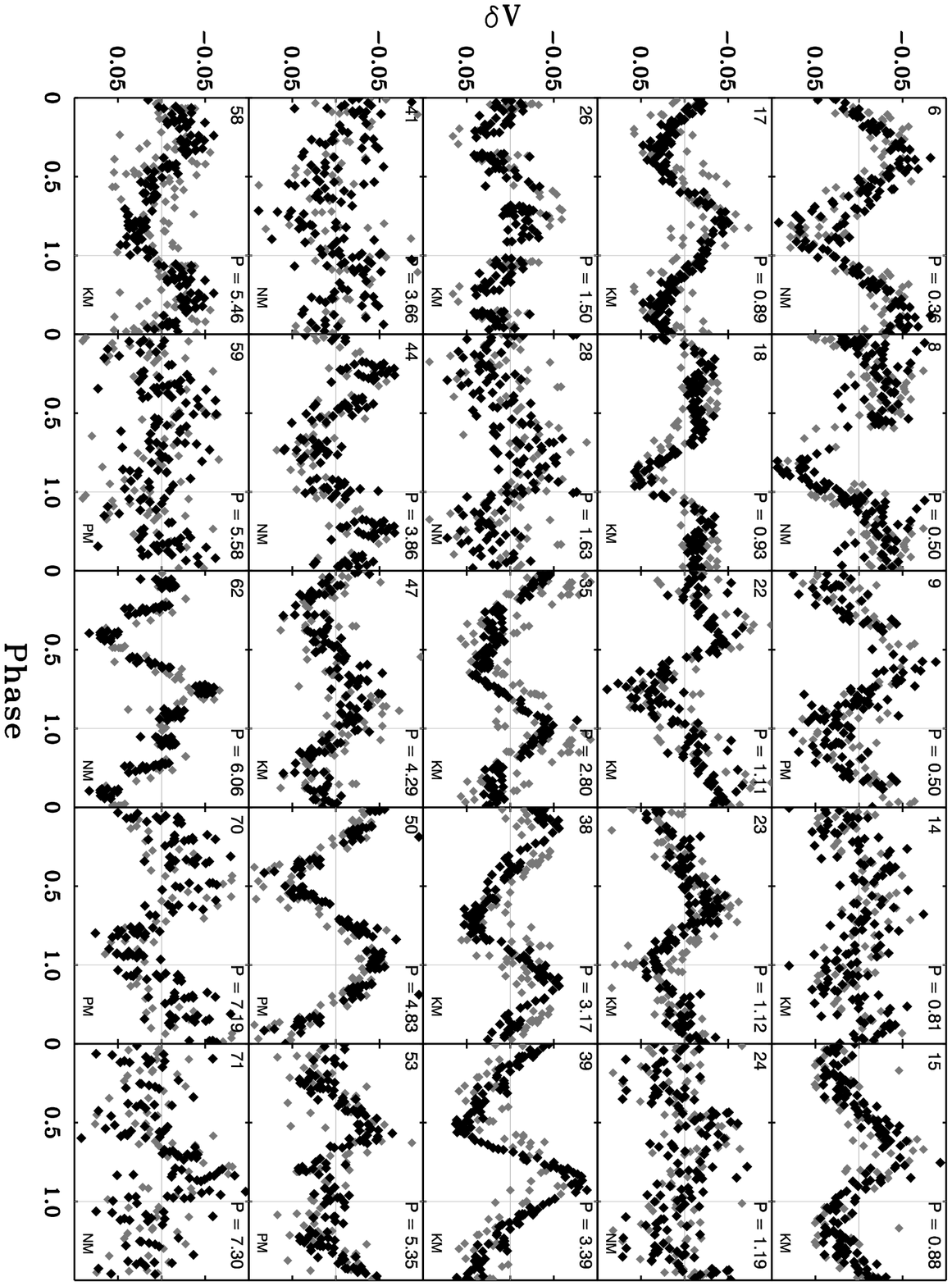}}\\
\centerline{Fig. 14. --- Continued.}
\clearpage
{\plotone{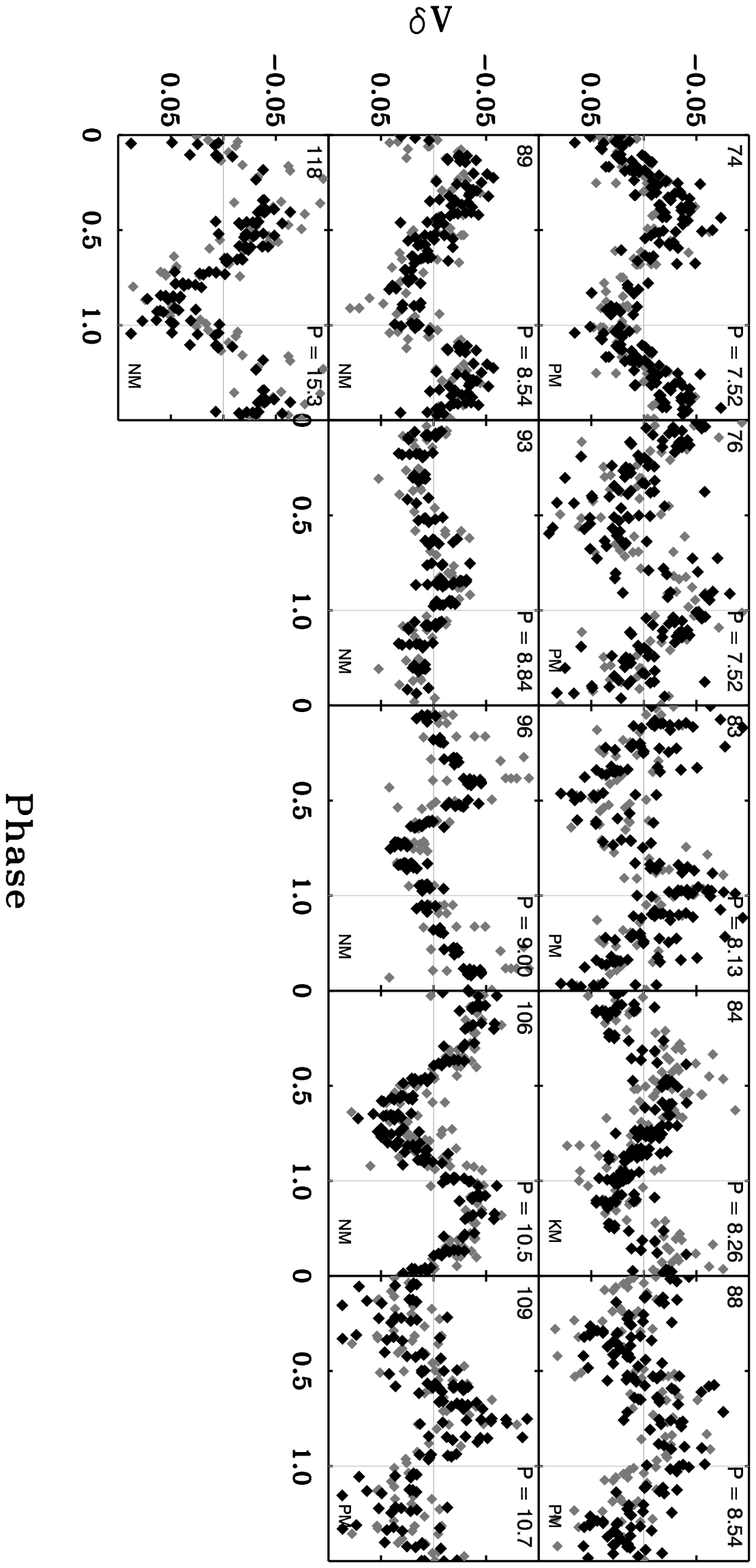}}\\
\centerline{Fig. 14. --- Continued.}
\clearpage
{\plotone{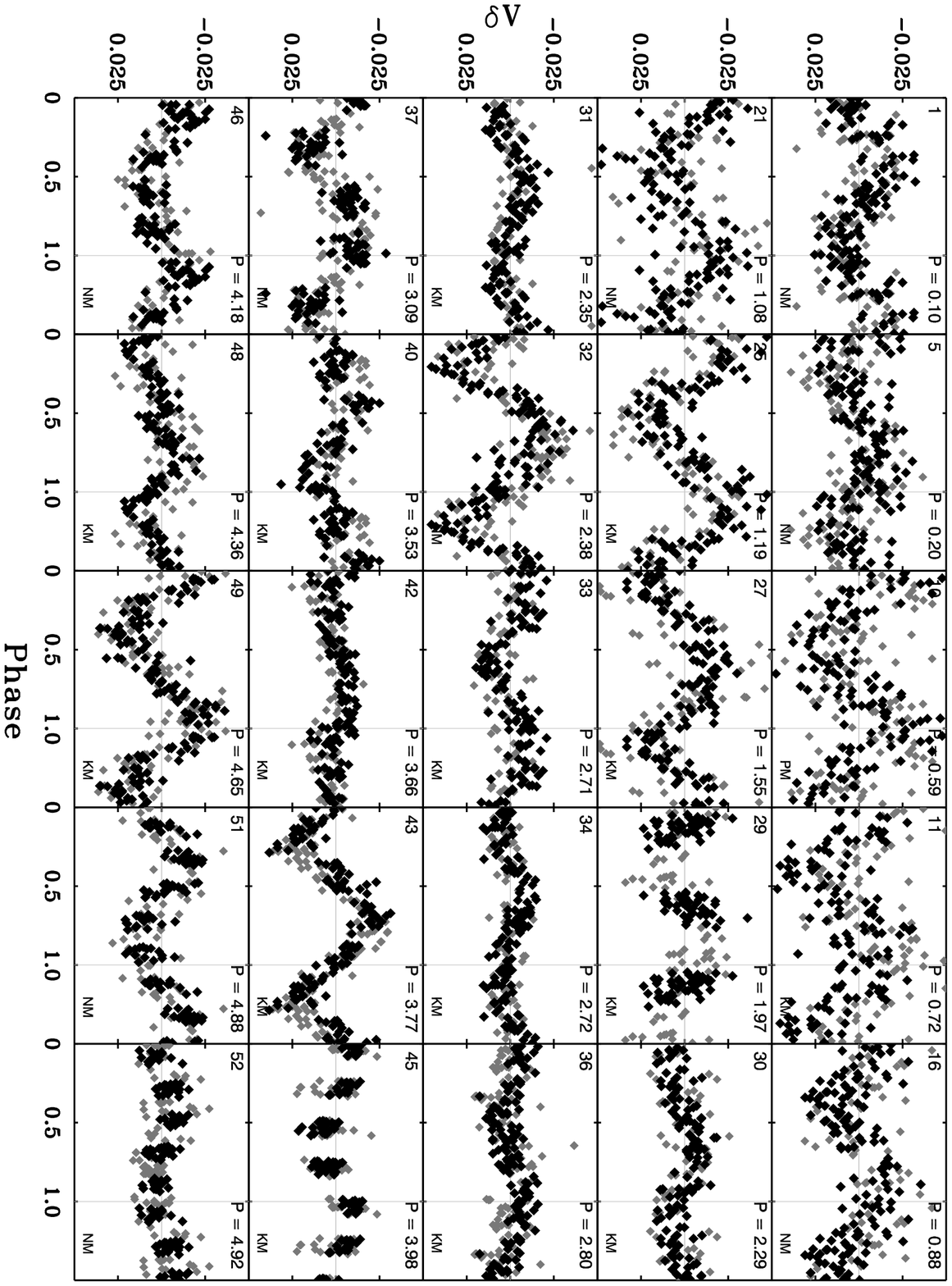}}\\
\centerline{Fig. 14. --- Continued.}
\clearpage
{\plotone{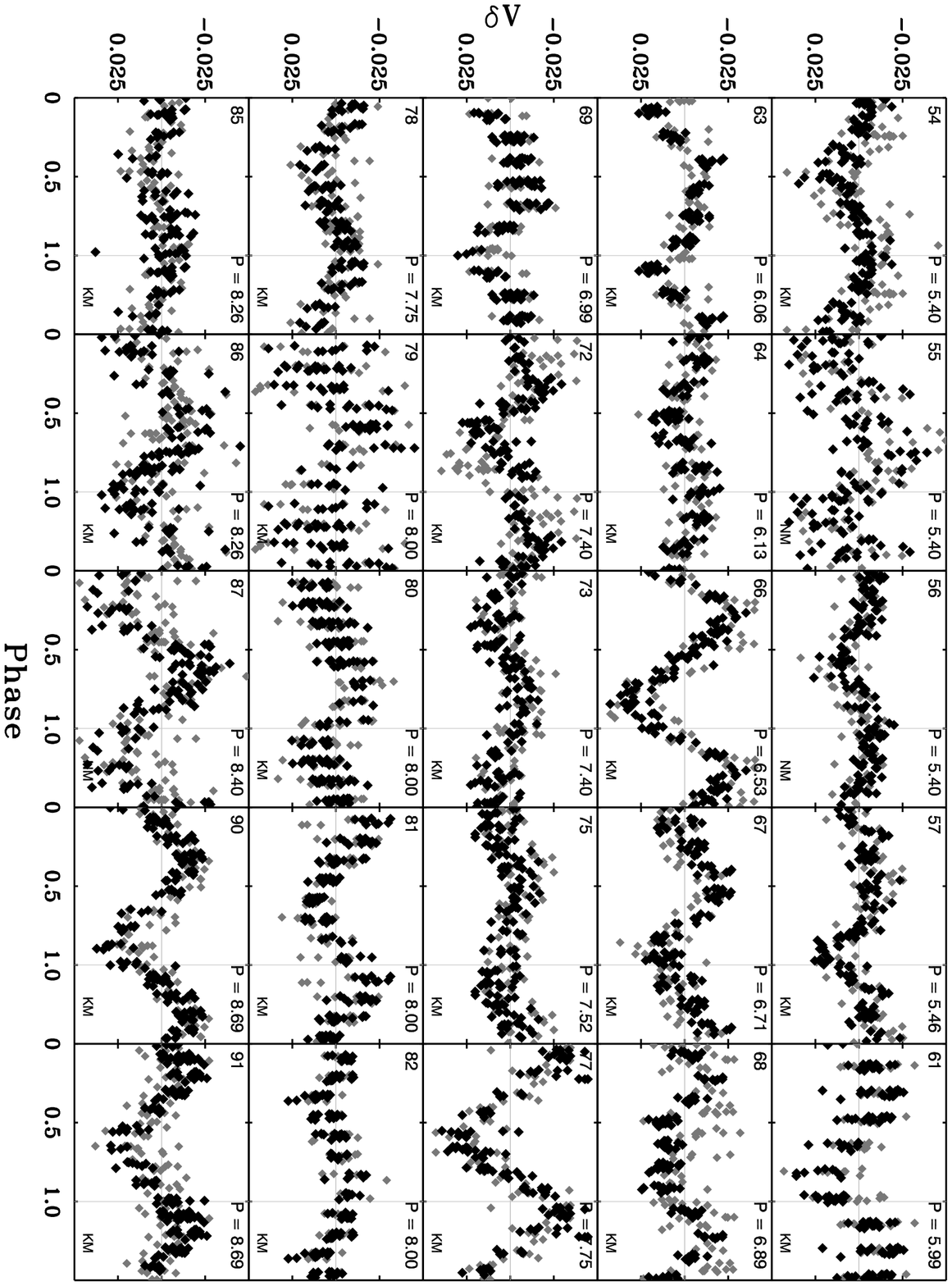}}\\
\centerline{Fig. 14. --- Continued.}
\clearpage
{\plotone{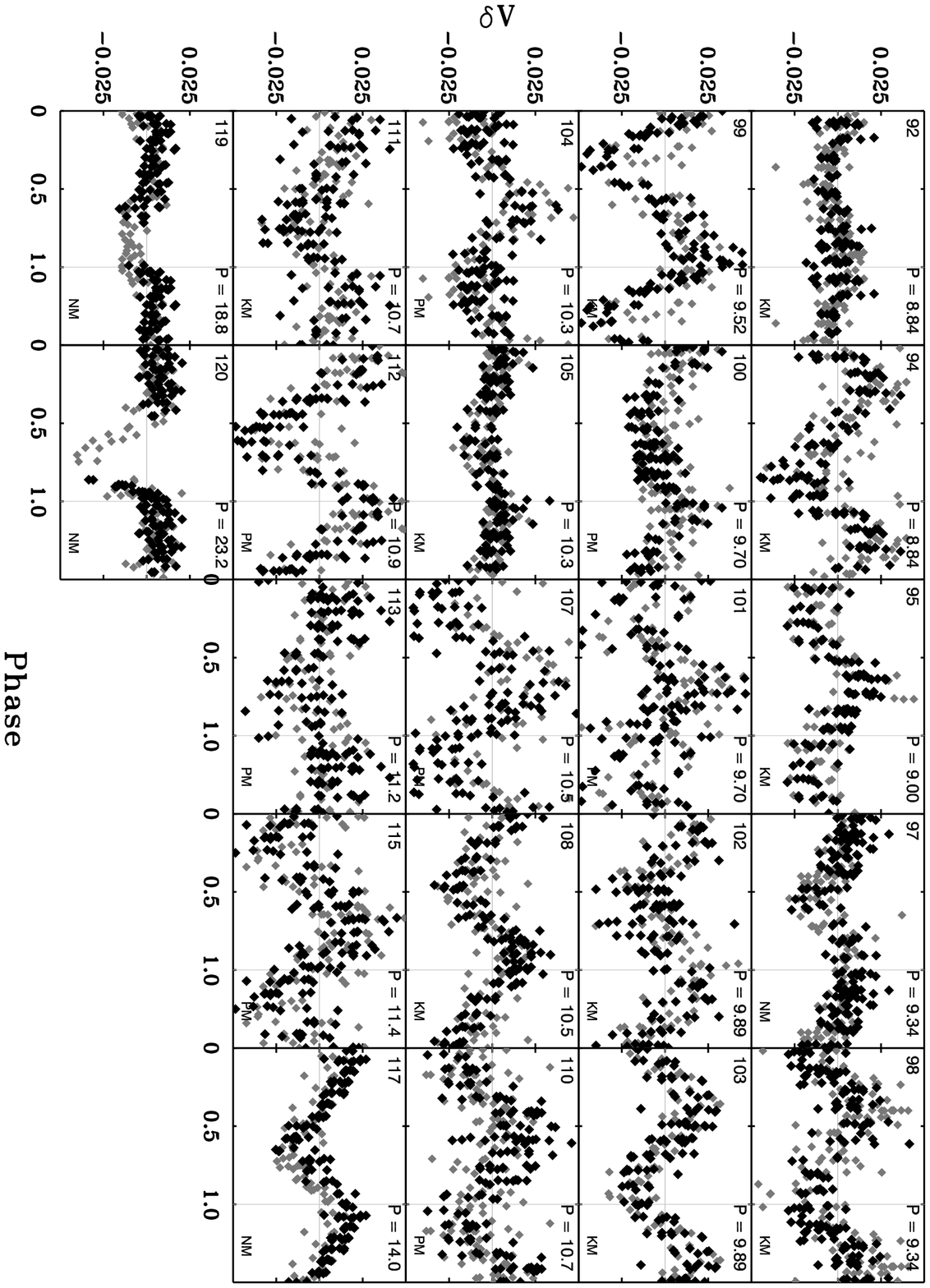}}\\
\centerline{Fig. 14. --- Continued.}
\clearpage
\pagestyle{plaintop}


\section{DATA FOR THE 120 STARS WITH MEASURED ROTATION PERIODS IN THE
FIELD OF M34}

Table~\ref{tab2} presents the results from this study together with
information relevant to this paper for 120 stars in the field of M34.
In the printed journal a stub version of Table 2 show the form of the
full table and a sample the first 5 lines of its contents. The full
version of the table can be found in the electronic edition of the
Journal. The stars appear in order of increasing rotation period and
the running number in the first column corresponds to the number in
the upper left hand corner of each light curve in Appendix A.
Columns 2 and 3 give the stellar equatorial coordinates (equinox 2000).
Column 4 lists the measured stellar rotation period in decimal days.
Columns 5 and 6 gives the stellar V magnitude and B-V color index,
respectively, corrected for extinction and reddening. Column
7 presents the number of radial-velocity measurements for the star and
columns 8 and 9 give the mean radial-velocity and the velocity standard
deviation, respectively. Column 10 list the radial-velocity cluster
membership probability calculated using the formalism by \citet{vkp58}.
Column 11 contains a proper-motion cluster membership probability from
\citet{jp96}. In column 12 we give the membership codes (mcode) also
found in the light curve plots as explained in Appendix A. Finally, in column
13 the rotational state of the star is indicated by a 1-letter code
representing, respectively, the $I$ sequence (``i''), the $C$ sequence (``c''),
and the gap (``g''). Stars with a ``-'' in column 13 have locations in
the color-period diagram that do not correspond to either of the sequences
or the gap.

\notetoeditor{A stub version of Table 2 with the first 5 lines can go here.
tab2_trun.tex contains just the first 5 lines of data. tab2.tex is
the full table. NOTE: in the "c", and "d" table notes, the first letter
is missing.}


\clearpage
\begin{deluxetable}{rccrrcccrrccccc}
\tabletypesize{\scriptsize}
\setlength{\tabcolsep}{0.8mm}
\tablecaption{Data for the 120 stars with measured rotation periods in the
field of M34
\label{tab2}}
\tablewidth{0pt}
\tablehead{
\colhead{No.}\tablenotemark{a} &
\colhead{RA} &
\colhead{DEC} &
\colhead{$P_{rot}$} &
\colhead{$V_{0}$} &
\colhead{$(B-V)_{0}$} &
\colhead{$N_{RV}$} &
\colhead{$\bar{RV}$} &
\colhead{$\sigma_{RV}$} &
\colhead{$P_{RV}$} &
\colhead{$P_{PM}$}\tablenotemark{b} &
\colhead{mcode}\tablenotemark{c} &
\colhead{Sequence}\tablenotemark{d} \\
\colhead{} &
\colhead{$h~~m~~s$} &
\colhead{$\degr~~'~~''$} &
\colhead{Days} &
\colhead{} &
\colhead{} &
\colhead{} &
\colhead{$km~s^{-1}$} &
\colhead{$km~s^{-1}$} &
\colhead{\%} &
\colhead{\%} &
\colhead{} &
\colhead{}
}
\startdata
     1 & 2 40 54.560 & 42 57 56.857 &   0.11 &  12.45 & 0.24   &   1 & 5.6    & 0.0   & 8   & ... & NM  & i   \\
     2 & 2 41 50.600 & 43 01 26.819 &   0.13 &  19.43 & ...    & ... & ...    & ...   & ... & ... & NM  & c   \\
     3 & 2 42 10.510 & 42 46 33.219 &   0.16 &  18.42 & 0.48   & ... & ...    & ...   & ... & ... & NM  & --  \\
     4 & 2 40 40.141 & 42 46 32.244 &   0.19 &  18.94 & 0.30   & ... & ...    & ...   & ... & ... & NM  & i   \\
     5 & 2 41 31.450 & 42 59 05.837 &   0.21 &  16.08 & 0.53   & ... & ...    & ...   & ... & ... & NM  & --  \\
     6 & 2 43 13.200 & 42 50 25.223 &   0.36 &  15.83 & 0.12   & ... & ...    & ...   & ... & ... & NM  & i   \\
     7 & 2 43 02.100 & 42 46 53.969 &   0.49 &  17.67 & 1.44   & ... & ...    & ...   & ... & ... & PM  & c   \\
     8 & 2 40 58.099 & 42 49 44.409 &   0.51 &  16.92 & 0.79   & ... & ...    & ...   & ... & ... & NM  & c   \\
     9 & 2 41 04.319 & 42 35 11.462 &   0.51 &  17.42 & 1.42   & ... & ...    & ...   & ... & ... & PM  & c   \\
    10 & 2 41 43.970 & 42 45 07.965 &   0.59 &  16.79 & 1.34   & ... & ...    & ...   & ... & ... & PM  & c   \\
    11 & 2 43 30.380 & 42 53 02.286 &   0.72 &  14.74 & 0.84   &  13 & -7.5   & 1.3   & 95  & 2   & KM  & c   \\
    12 & 2 43 00.049 & 42 58 01.224 &   0.79 &  15.60 & 1.04   &  12 & -13.4  & 3.0   & 6   & 47  & NM  & c   \\
    13 & 2 41 30.240 & 42 27 54.934 &   0.79 &  17.93 & 1.40   & ... & ...    & ...   & ... & ... & NM  & c   \\
    14 & 2 43 32.040 & 42 38 31.084 &   0.81 &  13.38 & 0.68   &   8 & -7.8   & 0.1   & 96  & 8   & KM  & c   \\
    15 & 2 41 53.320 & 42 35 25.800 &   0.88 &  16.08 & 1.16   &   6 & -6.9   & 1.7   & 91  & ... & KM  & c   \\
    16 & 2 42 17.510 & 42 35 03.566 &   0.88 &  15.53 & 0.29   & ... & ...    & ...   & ... & ... & NM  & i   \\
    17 & 2 41 35.330 & 42 41 01.995 &   0.90 &  14.33 & 0.79   &  15 & -7.7   & 1.8   & 95  & 86  & KM  & c   \\
    18 & 2 42 50.869 & 42 58 07.555 &   0.94 &  14.51 & 0.84   &  15 & -10.0  & 1.6   & 72  & 0   & KM  & c   \\
    19 & 2 42 06.570 & 42 39 36.755 &   0.96 &  17.66 & 1.41   & ... & ...    & ...   & ... & ... & NM  & c   \\
    20 & 2 42 31.619 & 42 37 09.950 &   1.07 &  15.97 & 1.13   &   9 & -9.9   & 1.8   & 74  & 46  & KM  & c   \\
    21 & 2 42 05.120 & 42 25 26.509 &   1.09 &  15.01 & 0.85   &  12 & -14.1  & 2.5   & 6   & ... & NM  & c   \\
    22 & 2 43 23.870 & 42 30 58.475 &   1.12 &  15.63 & 1.04   &  12 & -6.9   & 1.6   & 91  & ... & KM  & c   \\
    23 & 2 41 57.581 & 43 00 26.353 &   1.13 &  15.29 & 0.99   &  11 & -9.9   & 2.5   & 76  & 53  & KM  & c   \\
    24 & 2 42 26.029 & 42 48 53.871 &   1.19 &  17.32 & 0.93   & ... & ...    & ...   & ... & ... & NM  & c   \\
    25 & 2 42 47.650 & 42 45 46.596 &   1.19 &  16.45 & 1.20   &  11 & -7.2   & 1.3   & 94  & ... & KM  & c   \\
    26 & 2 42 49.180 & 42 40 13.875 &   1.50 &  15.73 & 1.03   &   5 & -7.7   & 0.3   & 95  & ... & KM  & c   \\
    27 & 2 42 57.940 & 42 58 03.559 &   1.55 &  14.78 & 0.86   &  13 & -6.3   & 1.7   & 81  & 0   & KM  & c   \\
    28 & 2 42 47.890 & 43 00 47.186 &   1.63 &  17.52 & 1.12   & ... & ...    & ...   & ... & ... & NM  & c   \\
    29 & 2 42 42.530 & 42 40 30.835 &   1.97 &  12.62 & 0.52   &  10 & -9.8   & 0.8   & 77  & 99  & KM  & i   \\
    30 & 2 41 33.661 & 42 32 21.985 &   2.29 &  11.99 & 0.51   &  11 & -8.4   & 0.9   & 95  & 99  & KM  & i   \\
    31 & 2 42 34.410 & 42 36 21.679 &   2.36 &  12.15 & 0.55   &  10 & -7.0   & 0.6   & 93  & 98  & KM  & i   \\
    32 & 2 41 58.850 & 42 57 30.105 &   2.39 &  15.60 & 0.83   & ... & ...    & ...   & ... & ... & NM  & g   \\
    33 & 2 42 08.270 & 42 38 41.535 &   2.71 &  12.90 & 0.54   &   3 & -7.1   & 0.2   & 93  & 97  & KM  & i   \\
    34 & 2 41 48.100 & 42 39 03.343 &   2.73 &  12.68 & 0.53   &  11 & -7.5   & 0.8   & 95  & 99  & KM  & i   \\
    35 & 2 41 36.720 & 42 50 20.018 &   2.80 &  14.44 & 0.82   &  11 & -7.6   & 0.8   & 95  & 44  & KM  & g   \\
    36 & 2 41 40.911 & 42 54 46.917 &   2.80 &  12.96 & 0.58   &   9 & -8.6   & 0.5   & 95  & 99  & KM  & i   \\
    37 & 2 42 15.010 & 42 54 33.376 &   3.10 &  13.04 & 0.62   &   8 & -2.4   & 0.2   & 7   & 0   & NM  & i   \\
    38 & 2 42 36.410 & 42 54 31.317 &   3.18 &  14.73 & 0.86   &  11 & -7.3   & 0.5   & 94  & 0   & KM  & g   \\
    39 & 2 41 48.760 & 42 27 58.477 &   3.39 &  14.70 & 0.96   &  15 & -8.3   & 5.5   & 95  & ... & KM  & g   \\
    40 & 2 43 15.420 & 42 30 36.186 &   3.54 &  12.90 & 0.54   &  15 & -8.5   & 1.7   & 95  & ... & KM  & i   \\
    41 & 2 42 44.011 & 42 26 39.733 &   3.67 &  17.59 & 0.89   & ... & ...    & ...   & ... & ... & NM  & g   \\
    42 & 2 42 51.481 & 42 47 52.375 &   3.67 &  12.78 & 0.64   &  15 & -8.2   & 1.1   & 95  & 99  & KM  & i   \\
    43 & 2 41 05.060 & 42 46 51.566 &   3.78 &  13.29 & 0.60   &   8 & -7.4   & 0.2   & 95  & 4   & KM  & i   \\
    44 & 2 42 57.960 & 42 41 46.119 &   3.86 &  16.13 & 1.36   & ... & ...    & ...   & ... & ... & NM  & g   \\
    45 & 2 41 26.390 & 42 48 12.934 &   3.99 &  13.21 & 0.59   &  10 & -8.2   & 0.3   & 96  & 98  & KM  & i   \\
    46 & 2 41 59.641 & 42 41 01.555 &   4.19 &  12.98 & 0.67   &  18 & -6.5   & 12.2  & 87  & 87  & KM  & i   \\
    47 & 2 43 18.140 & 42 46 36.968 &   4.29 &  13.76 & 0.69   &  16 & -8.1   & 28.3  & 96  & 94  & KM  & i   \\
    48 & 2 41 38.540 & 42 39 03.906 &   4.37 &  13.36 & 0.63   &   6 & -7.9   & 0.2   & 96  & 95  & KM  & i   \\
    49 & 2 42 14.080 & 42 55 47.974 &   4.65 &  13.89 & 0.77   &  10 & -7.8   & 0.4   & 96  & 17  & KM  & i   \\
    50 & 2 41 49.060 & 42 39 59.290 &   4.83 &  15.82 & 1.11   & ... & ...    & ...   & ... & ... & PM  & g   \\
    51 & 2 42 23.200 & 42 27 28.444 &   4.88 &  14.65 & 0.68   & ... & ...    & ...   & ... & ... & NM  & i   \\
    52 & 2 42 00.530 & 42 48 53.350 &   4.93 &  12.65 & 0.73   &  18 & -17.2  & 8.3   & 6   & 99  & NM  & i   \\
    53 & 2 41 51.771 & 42 38 22.913 &   5.35 &  16.57 & 1.28   & ... & ...    & ...   & ... & ... & PM  & g   \\
    54 & 2 41 37.561 & 42 59 39.414 &   5.41 &  13.86 & 0.72   &  12 & -9.4   & 0.6   & 88  & 3   & KM  & i   \\
    55 & 2 42 18.811 & 42 28 58.422 &   5.41 &  16.72 & 1.07   & ... & ...    & ...   & ... & ... & NM  & g   \\
    56 & 2 42 41.849 & 43 00 49.287 &   5.41 &  12.89 & 0.61   &  13 & -0.5   & 0.4   & 7   & 0   & NM  & i   \\
    57 & 2 41 33.510 & 42 42 11.360 &   5.47 &  13.74 & 0.70   &   7 & -8.9   & 0.7   & 93  & 5   & KM  & i   \\
    58 & 2 41 48.579 & 42 49 33.532 &   5.47 &  16.02 & 1.07   &  10 & -0.6   & 1.3   & 7   & 17  & NM  & g   \\
    59 & 2 41 36.680 & 42 40 03.438 &   5.59 &  17.91 & 1.49   & ... & ...    & ...   & ... & ... & PM  & g   \\
    60 & 2 41 32.840 & 43 02 16.148 &   5.99 &  17.74 & 1.52   & ... & ...    & ...   & ... & ... & PM  & g   \\
    61 & 2 41 38.240 & 42 44 04.231 &   5.99 &  13.87 & 0.73   &  10 & -7.9   & 0.3   & 96  & 93  & KM  & i   \\
    62 & 2 40 44.530 & 42 54 04.373 &   6.06 &  12.64 & 0.98   &   1 & -55.8  & 0.0   & 15  & ... & NM  & g   \\
    63 & 2 41 47.970 & 42 41 18.076 &   6.06 &  11.97 & 0.55   &  10 & -8.4   & 0.4   & 95  & 0   & KM  & --  \\
    64 & 2 42 06.660 & 42 36 20.484 &   6.14 &  13.49 & 0.80   &   4 & -6.8   & 0.1   & 90  & 95  & KM  & i   \\
    65 & 2 42 10.400 & 42 59 35.885 &   6.37 &  17.77 & 1.50   & ... & ...    & ...   & ... & ... & PM  & g   \\
    66 & 2 42 56.980 & 42 35 21.281 &   6.54 &  14.19 & 0.80   &   7 & -8.2   & 0.2   & 96  & 0   & KM  & i   \\
    67 & 2 41 49.430 & 42 36 37.046 &   6.71 &  14.19 & 0.81   &   9 & -7.3   & 0.4   & 94  & 91  & KM  & i   \\
    68 & 2 42 11.430 & 42 50 36.346 &   6.90 &  14.27 & 0.85   &   8 & -8.1   & 0.3   & 96  & 0   & KM  & i   \\
    69 & 2 41 05.251 & 42 56 43.029 &   7.00 &  14.37 & 0.81   &   8 & -8.1   & 0.3   & 96  & ... & KM  & i   \\
    70 & 2 40 42.850 & 42 38 58.660 &   7.20 &  16.42 & 1.27   & ... & ...    & ...   & ... & ... & PM  & g   \\
    71 & 2 42 50.400 & 42 34 19.882 &   7.30 &  17.76 & 1.41   & ... & ...    & ...   & ... & ... & NM  & g   \\
    72 & 2 41 01.090 & 42 52 46.521 &   7.41 &  14.70 & 0.87   &  13 & -7.5   & 0.9   & 95  & 0   & KM  & i   \\
    73 & 2 43 10.360 & 42 54 32.621 &   7.41 &  14.57 & 0.90   &   5 & -6.7   & 0.4   & 89  & 72  & KM  & i   \\
    74 & 2 40 45.000 & 42 53 47.014 &   7.52 &  16.60 & 1.28   & ... & ...    & ...   & ... & ... & PM  & g   \\
    75 & 2 41 35.160 & 42 33 30.539 &   7.52 &  13.92 & 0.86   &  10 & -7.9   & 0.5   & 96  & 0   & KM  & i   \\
    76 & 2 42 11.080 & 42 43 16.083 &   7.52 &  16.49 & 1.29   & ... & ...    & ...   & ... & ... & PM  & g   \\
    77 & 2 41 23.170 & 42 40 15.523 &   7.75 &  14.74 & 0.90   &  12 & -7.1   & 0.5   & 93  & 84  & KM  & i   \\
    78 & 2 42 47.840 & 42 47 42.790 &   7.75 &  14.73 & 0.90   &  11 & -8.8   & 0.6   & 94  & 67  & KM  & i   \\
    79 & 2 41 01.950 & 42 42 31.163 &   8.00 &  14.85 & 0.91   &   9 & -7.5   & 0.4   & 95  & 76  & KM  & i   \\
    80 & 2 41 06.299 & 42 46 21.010 &   8.00 &  15.41 & 1.03   &  15 & -7.3   & 5.9   & 94  & 26  & KM  & i   \\
    81 & 2 41 51.379 & 42 34 24.372 &   8.00 &  14.72 & 0.89   &   8 & -8.9   & 0.4   & 93  & 0   & KM  & i   \\
    82 & 2 42 26.400 & 42 30 13.074 &   8.00 &  14.48 & 0.84   &  12 & -7.1   & 0.7   & 94  & 85  & KM  & i   \\
    83 & 2 42 04.990 & 42 37 46.905 &   8.13 &  17.58 & 1.52   & ... & ...    & ...   & ... & ... & PM  & g   \\
    84 & 2 40 49.760 & 42 46 54.999 &   8.27 &  15.52 & 1.05   &  11 & -8.7   & 0.8   & 94  & ... & KM  & i   \\
    85 & 2 41 21.480 & 42 35 43.996 &   8.27 &  14.98 & 0.94   &  11 & -9.0   & 0.4   & 92  & 10  & KM  & i   \\
    86 & 2 42 05.880 & 42 46 53.475 &   8.27 &  14.34 & 0.99   &   9 & -8.5   & 0.5   & 95  & 71  & KM  & i   \\
    87 & 2 42 02.370 & 43 01 13.320 &   8.40 &  14.73 & 0.98   &  15 & -2.8   & 0.6   & 7   & ... & NM  & i   \\
    88 & 2 41 44.269 & 42 35 35.454 &   8.55 &  17.26 & 1.38   & ... & ...    & ...   & ... & ... & PM  & g   \\
    89 & 2 43 03.560 & 43 03 01.563 &   8.55 &  15.61 & ...    & ... & ...    & ...   & ... & ... & NM  & g   \\
    90 & 2 41 30.691 & 42 28 53.423 &   8.70 &  14.80 & 0.89   &  11 & -7.7   & 0.6   & 95  & ... & KM  & i   \\
    91 & 2 42 02.650 & 42 51 51.411 &   8.70 &  15.02 & 0.95   &  12 & -7.4   & 0.6   & 95  & 76  & KM  & i   \\
    92 & 2 41 24.320 & 42 48 18.550 &   8.85 &  14.71 & 1.01   &  10 & -6.9   & 0.6   & 92  & 0   & KM  & i   \\
    93 & 2 41 27.290 & 42 43 41.970 &   8.85 &  14.89 & 0.83   &   9 & -24.4  & 0.4   & 6   & 0   & NM  & i   \\
    94 & 2 43 21.230 & 42 48 56.069 &   8.85 &  15.74 & 1.18   &  13 & -10.5  & 0.8   & 44  & 24  & KM  & i   \\
    95 & 2 41 42.990 & 42 33 13.277 &   9.01 &  15.43 & 1.09   &  13 & -9.3   & 0.5   & 89  & 5   & KM  & i   \\
    96 & 2 42 25.920 & 42 28 44.057 &   9.01 &  13.80 & 1.21   & ... & ...    & ...   & ... & ... & NM  & i   \\
    97 & 2 41 13.010 & 42 50 19.112 &   9.35 &  15.71 & 0.89   & ... & ...    & ...   & ... & ... & NM  & i   \\
    98 & 2 41 56.050 & 42 58 30.558 &   9.35 &  15.90 & 1.18   &   9 & -8.3   & 0.7   & 95  & ... & KM  & i   \\
    99 & 2 42 24.871 & 42 27 39.595 &   9.52 &  15.84 & 1.14   &  12 & -8.1   & 0.7   & 96  & ... & KM  & i   \\
   100 & 2 41 37.671 & 42 57 21.426 &   9.71 &  15.45 & 1.05   & ... & ...    & ...   & ... & ... & PM  & i   \\
   101 & 2 42 25.010 & 42 53 25.824 &   9.71 &  17.18 & 1.42   & ... & ...    & ...   & ... & ... & PM  & g   \\
   102 & 2 41 19.321 & 42 35 28.148 &   9.90 &  16.32 & 1.28   &   8 & -8.0   & 1.1   & 96  & ... & KM  & i   \\
   103 & 2 42 59.660 & 42 33 13.991 &   9.90 &  15.45 & 1.06   &  14 & -7.6   & 44.8  & 95  & 6   & KM  & i   \\
   104 & 2 41 46.440 & 42 32 31.501 &  10.31 &  16.12 & 1.20   & ... & ...    & ...   & ... & ... & PM  & i   \\
   105 & 2 42 55.190 & 42 45 50.249 &  10.31 &  15.69 & 1.13   &   9 & -7.2   & 0.8   & 94  & 27  & KM  & i   \\
   106 & 2 41 38.340 & 42 55 42.865 &  10.52 &  15.73 & 0.91   & ... & ...    & ...   & ... & ... & NM  & i   \\
   107 & 2 42 11.749 & 42 31 17.440 &  10.52 &  16.70 & 1.31   & ... & ...    & ...   & ... & ... & PM  & i   \\
   108 & 2 42 59.650 & 42 48 23.426 &  10.52 &  15.49 & 1.25   &  10 & -6.7   & 0.9   & 90  & 43  & KM  & i   \\
   109 & 2 41 38.931 & 42 43 00.963 &  10.75 &  17.44 & 1.43   & ... & ...    & ...   & ... & ... & PM  & i   \\
   110 & 2 41 45.020 & 42 53 56.215 &  10.75 &  16.12 & 1.37   & ... & ...    & ...   & ... & ... & PM  & i   \\
   111 & 2 42 43.550 & 42 52 30.027 &  10.75 &  15.59 & 1.18   &  12 & -10.4  & 0.8   & 49  & 19  & KM  & i   \\
   112 & 2 42 20.410 & 42 49 05.517 &  10.99 &  16.31 & 1.26   & ... & ...    & ...   & ... & ... & PM  & i   \\
   113 & 2 41 01.740 & 42 37 34.573 &  11.23 &  16.23 & 1.22   & ... & ...    & ...   & ... & ... & PM  & i   \\
   114 & 2 41 31.200 & 42 41 13.943 &  11.23 &  17.91 & 1.48   & ... & ...    & ...   & ... & ... & NM  & i   \\
   115 & 2 40 49.190 & 42 48 20.432 &  11.49 &  16.39 & 1.28   & ... & ...    & ...   & ... & ... & PM  & i   \\
   116 & 2 40 38.520 & 42 43 33.826 &  12.98 &  18.50 & 0.50   & ... & ...    & ...   & ... & ... & NM  & i   \\
   117 & 2 42 23.059 & 42 57 28.814 &  14.08 &  14.37 & 0.79   &   8 & -28.0  & 1.4   & 6   & 0   & NM  & i   \\
   118 & 2 43 13.039 & 42 32 29.496 &  15.37 &  16.79 & 0.93   & ... & ...    & ...   & ... & ... & NM  & i   \\
   119 & 2 41 23.820 & 42 30 55.811 &  18.85 &  14.62 & 0.94   &   5 & 6.0    & 0.2   & 8   & 0   & NM  & i   \\
   120 & 2 43 07.390 & 42 47 46.772 &  23.22 &  12.36 & 1.25   &   1 & -6.8   & 0.0   & 90  & 0   & NM  & i   \\
\enddata
\tablecomments{This stub version of Table 2 is intended to show the form of the full table and a sample of its contents. The full version of the table can be found online.}

\tablenotetext{a}{Running number assigned to 120 rotators after sorting by rotation period.}
\tablenotetext{b}{Proper-motion membership probabilities are from \citep{jp96}}
\tablenotetext{c}{Letter code denoting a star's cluster membership status (see introductory text of Appendix for code meaning).}
\tablenotetext{d}{Letters ``i'', ``c'', and ``g'' mark stars on the I and C sequances or in the gap, respectively.}

\end{deluxetable}

\clearpage



\begin{thebibliography}{85}
\expandafter\ifx\csname natexlab\endcsname\relax\def\natexlab#1{#1}\fi

\bibitem[{{Barnes} \& {Sofia}(1996)}]{bs96}
{Barnes}, S., \& {Sofia}, S. 1996, \apj, 462, 746

\bibitem[{{Barnes} {et~al.}(2001){Barnes}, {Sofia}, \& {Pinsonneault}}]{bsp01}
{Barnes}, S., {Sofia}, S., \& {Pinsonneault}, M. 2001, \apj, 548, 1071

\bibitem[{{Barnes}(2003)}]{barnes03a}
{Barnes}, S.~A. 2003, \apj, 586, 464

\bibitem[{{Barnes}(2007)}]{barnes07}
---. 2007, \apj, 669, 1167

\bibitem[{{Barnes} {et~al.}(1999){Barnes}, {Sofia}, {Prosser}, \&
  {Stauffer}}]{bsp+99}
{Barnes}, S.~A., {Sofia}, S., {Prosser}, C.~F., \& {Stauffer}, J.~R. 1999,
  \apj, 516, 263

\bibitem[{{Berdyugina} \& {J{\"a}rvinen}(2005)}]{bj05}
{Berdyugina}, S.~V., \& {J{\"a}rvinen}, S.~P. 2005, Astronomische Nachrichten,
  326, 283

\bibitem[{{Berdyugina} \& {Usoskin}(2003)}]{bu03}
{Berdyugina}, S.~V., \& {Usoskin}, I.~G. 2003, \aap, 405, 1121

\bibitem[{{Bogdan} {et~al.}(1988){Bogdan}, {Gilman}, {Lerche}, \&
  {Howard}}]{bgl+88}
{Bogdan}, T.~J., {Gilman}, P.~A., {Lerche}, I., \& {Howard}, R. 1988, \apj,
  327, 451

\bibitem[{{Bouvier} {et~al.}(1997){Bouvier}, {Forestini}, \& {Allain}}]{bfa97}
{Bouvier}, J., {Forestini}, M., \& {Allain}, S. 1997, \aap, 326, 1023

\bibitem[{{Caldwell} {et~al.}(1993){Caldwell}, {Cousins}, {Ahlers}, {van
  Wamelen}, \& {Maritz}}]{cca+93}
{Caldwell}, J.~A.~R., {Cousins}, A.~W.~J., {Ahlers}, C.~C., {van Wamelen}, P.,
  \& {Maritz}, E.~J. 1993, South African Astronomical Observatory Circular, 15,
  1

\bibitem[{{Canterna} {et~al.}(1979){Canterna}, {Crawford}, \& {Perry}}]{ccp79}
{Canterna}, R., {Crawford}, D.~L., \& {Perry}, C.~L. 1979, \pasp, 91, 263

\bibitem[{{Chaboyer} {et~al.}(1995){Chaboyer}, {Demarque}, \&
  {Pinsonneault}}]{cdp95}
{Chaboyer}, B., {Demarque}, P., \& {Pinsonneault}, M.~H. 1995, \apj, 441, 865

\bibitem[{{Collier Cameron} {et~al.}(2009){Collier Cameron}, {Davidson},
  {Hebb}, {Skinner}, {Anderson}, {Christian}, {Clarkson}, {Enoch}, {Irwin},
  {Joshi}, {Haswell}, {Hellier}, {Horne}, {Kane}, {Lister}, {Maxted}, {Norton},
  {Parley}, {Pollacco}, {Ryans}, {Scholz}, {Skillen}, {Smalley}, {Street},
  {West}, {Wilson}, \& {Wheatley}}]{cdh+09}
{Collier Cameron}, A., {Davidson}, V.~A., {Hebb}, L., {Skinner}, G.,
  {Anderson}, D.~R., {Christian}, D.~J., {Clarkson}, W.~I., {Enoch}, B.,
  {Irwin}, J., {Joshi}, Y., {Haswell}, C.~A., {Hellier}, C., {Horne}, K.~D.,
  {Kane}, S.~R., {Lister}, T.~A., {Maxted}, P.~F.~L., {Norton}, A.~J.,
  {Parley}, N., {Pollacco}, D., {Ryans}, R., {Scholz}, A., {Skillen}, I.,
  {Smalley}, B., {Street}, R.~A., {West}, R.~G., {Wilson}, D.~M., \&
  {Wheatley}, P.~J. 2009, \mnras, 1358

\bibitem[{{Deliyannis}(2010)}]{deliyannis10}
{Deliyannis}, C.~P. 2010, private comm.

\bibitem[{{Denissenkov} {et~al.}(2010){Denissenkov}, {Pinsonneault},
  {Terndrup}, \& {Newsham}}]{dpt+10}
{Denissenkov}, P.~A., {Pinsonneault}, M., {Terndrup}, D.~M., \& {Newsham}, G.
  2010, \apj, 716, 1269

\bibitem[{{Donahue} {et~al.}(1996){Donahue}, {Saar}, \& {Baliunas}}]{dsb96}
{Donahue}, R.~A., {Saar}, S.~H., \& {Baliunas}, S.~L. 1996, \apj, 466, 384

\bibitem[{{Endal} \& {Sofia}(1981)}]{es81}
{Endal}, A.~S., \& {Sofia}, S. 1981, \apj, 243, 625

\bibitem[{{Geller} {et~al.}(2008){Geller}, {Mathieu}, {Harris}, \&
  {McClure}}]{gmh+08}
{Geller}, A.~M., {Mathieu}, R.~D., {Harris}, H.~C., \& {McClure}, R.~D. 2008,
  \aj, 135, 2264

\bibitem[{{Hartman} {et~al.}(2010{\natexlab{a}}){Hartman}, {Bakos},
  {Kov{\'a}cs}, \& {Noyes}}]{hbk+10}
{Hartman}, J.~D., {Bakos}, G.~{\'A}., {Kov{\'a}cs}, G., \& {Noyes}, R.~W.
  2010{\natexlab{a}}, ArXiv e-prints

\bibitem[{{Hartman} {et~al.}(2010{\natexlab{b}}){Hartman}, {Bakos}, \&
  {Pal}}]{hbn+10}
{Hartman}, J.~D., {Bakos}, G.~{\'A.}, N. R.~W. S. B. M. T. S.~A., \& {Pal}, A.
  2010{\natexlab{b}}, ArXiv e-prints

\bibitem[{{Hartman} {et~al.}(2009){Hartman}, {Gaudi}, {Pinsonneault}, {Stanek},
  {Holman}, {McLeod}, {Meibom}, {Barranco}, \& {Kalirai}}]{hgp+09}
{Hartman}, J.~D., {Gaudi}, B.~S., {Pinsonneault}, M.~H., {Stanek}, K.~Z.,
  {Holman}, M.~J., {McLeod}, B.~A., {Meibom}, S., {Barranco}, J.~A., \&
  {Kalirai}, J.~S. 2009, \apj, 691, 342

\bibitem[{{Herbst} {et~al.}(2007){Herbst}, {Eisl{\"o}ffel}, {Mundt}, \&
  {Scholz}}]{hem+07}
{Herbst}, W., {Eisl{\"o}ffel}, J., {Mundt}, R., \& {Scholz}, A. 2007, in
  Protostars and Planets V, ed. B.~{Reipurth}, D.~{Jewitt}, \& K.~{Keil},
  297--311

\bibitem[{{Honeycutt}(1992)}]{honeycutt92}
{Honeycutt}, R.~K. 1992, \pasp, 104, 435

\bibitem[{{Hussain}(2002)}]{hussain02}
{Hussain}, G.~A.~J. 2002, Astronomische Nachrichten, 323, 349

\bibitem[{{Ianna} \& {Schlemmer}(1993)}]{is93}
{Ianna}, P.~A., \& {Schlemmer}, D.~M. 1993, \aj, 105, 209

\bibitem[{{Irwin} {et~al.}(2009){Irwin}, {Aigrain}, {Bouvier}, {Hebb},
  {Hodgkin}, {Irwin}, \& {Moraux}}]{iab+09}
{Irwin}, J., {Aigrain}, S., {Bouvier}, J., {Hebb}, L., {Hodgkin}, S., {Irwin},
  M., \& {Moraux}, E. 2009, \mnras, 392, 1456

\bibitem[{{Irwin} {et~al.}(2006){Irwin}, {Aigrain}, {Hodgkin}, {Irwin},
  {Bouvier}, {Clarke}, {Hebb}, \& {Moraux}}]{iah+06}
{Irwin}, J., {Aigrain}, S., {Hodgkin}, S., {Irwin}, M., {Bouvier}, J.,
  {Clarke}, C., {Hebb}, L., \& {Moraux}, E. 2006, \mnras, 370, 954

\bibitem[{{Irwin} {et~al.}(2010){Irwin}, {Berta}, {Burke}, {Charbonneau},
  {Nutzman}, {West}, \& {Falco}}]{ibb+10}
{Irwin}, J., {Berta}, Z.~K., {Burke}, C.~J., {Charbonneau}, D., {Nutzman}, P.,
  {West}, A.~A., \& {Falco}, E.~E. 2010, ArXiv e-prints

\bibitem[{{Irwin} {et~al.}(2007){Irwin}, {Hodgkin}, {Aigrain}, {Hebb},
  {Bouvier}, {Clarke}, {Moraux}, \& {Bramich}}]{iha+07}
{Irwin}, J., {Hodgkin}, S., {Aigrain}, S., {Hebb}, L., {Bouvier}, J., {Clarke},
  C., {Moraux}, E., \& {Bramich}, D.~M. 2007, \mnras, 377, 741

\bibitem[{{James} {et~al.}(2010){James}, {Barnes}, {Meibom}, {Lockwood},
  {Levine}, {Deliyannis}, {Platais}, {Steinhauer}, \& {Hurley}}]{jbm+10}
{James}, D.~J., {Barnes}, S.~A., {Meibom}, S., {Lockwood}, G.~W., {Levine},
  S.~E., {Deliyannis}, C., {Platais}, I., {Steinhauer}, A., \& {Hurley}, B.~K.
  2010, \aap, 515, A100+

\bibitem[{{Jeffers} {et~al.}(2006){Jeffers}, {Barnes}, {Cameron}, \&
  {Donati}}]{jbc+06}
{Jeffers}, S.~V., {Barnes}, J.~R., {Cameron}, A.~C., \& {Donati}, J. 2006,
  \mnras, 366, 667

\bibitem[{{Jianke} \& {Collier Cameron}(1993)}]{jc93}
{Jianke}, L., \& {Collier Cameron}, A. 1993, \mnras, 261, 766

\bibitem[{{Jones} \& {Prosser}(1996)}]{jp96}
{Jones}, B.~F., \& {Prosser}, C.~F. 1996, \aj, 111, 1193

\bibitem[{{Kalirai} {et~al.}(2003){Kalirai}, {Fahlman}, {Richer}, \&
  {Ventura}}]{kfr+03}
{Kalirai}, J.~S., {Fahlman}, G.~G., {Richer}, H.~B., \& {Ventura}, P. 2003,
  \aj, 126, 1402

\bibitem[{{Kawaler}(1988)}]{kawaler88}
{Kawaler}, S.~D. 1988, \apj, 333, 236

\bibitem[{{Koenigl}(1991)}]{konigl91}
{Koenigl}, A. 1991, \apjl, 370, L39

\bibitem[{{Korhonen} \& {J{\"a}rvinen}(2007)}]{kj07}
{Korhonen}, H., \& {J{\"a}rvinen}, S.~P. 2007, in IAU Symposium, Vol. 240, IAU
  Symposium, ed. {W.~I.~Hartkopf, E.~F.~Guinan, \& P.~Harmanec}, 453--455

\bibitem[{{Kraft}(1967)}]{kraft67}
{Kraft}, R.~P. 1967, \apj, 150, 551

\bibitem[{{Krishnamurthi} {et~al.}(1997){Krishnamurthi}, {Pinsonneault},
  {Barnes}, \& {Sofia}}]{kpb+97}
{Krishnamurthi}, A., {Pinsonneault}, M.~H., {Barnes}, S., \& {Sofia}, S. 1997,
  \apj, 480, 303

\bibitem[{{Krishnamurthi} {et~al.}(1998){Krishnamurthi}, {Terndrup},
  {Pinsonneault}, {Sellgren}, {Stauffer}, {Schild}, {Backman}, {Beisser},
  {Dahari}, {Dasgupta}, {Hagelgans}, {Seeds}, {Anand}, {Laaksonen},
  {Marschall}, \& {Ramseyer}}]{ktp+98}
{Krishnamurthi}, A., {Terndrup}, D.~M., {Pinsonneault}, M.~H., {Sellgren}, K.,
  {Stauffer}, J.~R., {Schild}, R., {Backman}, D.~E., {Beisser}, K.~B.,
  {Dahari}, D.~B., {Dasgupta}, A., {Hagelgans}, J.~T., {Seeds}, M.~A., {Anand},
  R., {Laaksonen}, B.~D., {Marschall}, L.~A., \& {Ramseyer}, T. 1998, \apj,
  493, 914

\bibitem[{{Lanza} {et~al.}(2009){Lanza}, {Pagano}, {Leto}, {Messina},
  {Aigrain}, {Alonso}, {Auvergne}, {Baglin}, {Barge}, {Bonomo}, {Boumier},
  {Collier Cameron}, {Comparato}, {Cutispoto}, {de Medeiros}, {Foing},
  {Kaiser}, {Moutou}, {Parihar}, {Silva-Valio}, \& {Weiss}}]{lpl+09}
{Lanza}, A.~F., {Pagano}, I., {Leto}, G., {Messina}, S., {Aigrain}, S.,
  {Alonso}, R., {Auvergne}, M., {Baglin}, A., {Barge}, P., {Bonomo}, A.~S.,
  {Boumier}, P., {Collier Cameron}, A., {Comparato}, M., {Cutispoto}, G., {de
  Medeiros}, J.~R., {Foing}, B., {Kaiser}, A., {Moutou}, C., {Parihar}, P.~S.,
  {Silva-Valio}, A., \& {Weiss}, W.~W. 2009, \aap, 493, 193

\bibitem[{{MacGregor} \& {Brenner}(1991)}]{mb91}
{MacGregor}, K.~B., \& {Brenner}, M. 1991, \apj, 376, 204

\bibitem[{{Mamajek} \& {Hillenbrand}(2008)}]{mh08}
{Mamajek}, E.~E., \& {Hillenbrand}, L.~A. 2008, \apj, 687, 1264

\bibitem[{{Mathieu}(2000)}]{mathieu00}
{Mathieu}, R.~D. 2000, in ASP Conf. Ser. 198: Stellar Clusters and
  Associations: Convection, Rotation, and Dynamos, 517

\bibitem[{{Meibom} {et~al.}(2006){Meibom}, {Mathieu}, \& {Stassun}}]{mms06}
{Meibom}, S., {Mathieu}, R.~D., \& {Stassun}, K.~G. 2006, \apj, 653, 621

\bibitem[{{Meibom} {et~al.}(2009){Meibom}, {Mathieu}, \& {Stassun}}]{mms09}
---. 2009, \apj, 695, 679

\bibitem[{{Meynet} {et~al.}(1993){Meynet}, {Mermilliod}, \& {Maeder}}]{mmm93}
{Meynet}, G., {Mermilliod}, J., \& {Maeder}, A. 1993, \aaps, 98, 477

\bibitem[{{Mosser} {et~al.}(2009){Mosser}, {Baudin}, {Lanza}, {Hulot},
  {Catala}, {Baglin}, \& {Auvergne}}]{mbl+09}
{Mosser}, B., {Baudin}, F., {Lanza}, A.~F., {Hulot}, J.~C., {Catala}, C.,
  {Baglin}, A., \& {Auvergne}, M. 2009, \aap, 506, 245

\bibitem[{{O'Neal} {et~al.}(1996){O'Neal}, {Saar}, \& {Neff}}]{osn96}
{O'Neal}, D., {Saar}, S.~H., \& {Neff}, J.~E. 1996, \apj, 463, 766

\bibitem[{{Patten} \& {Simon}(1996)}]{ps96}
{Patten}, B.~M., \& {Simon}, T. 1996, \apjs, 106, 489

\bibitem[{{Perryman} {et~al.}(1998){Perryman}, {Brown}, {Lebreton}, {Gomez},
  {Turon}, {de Strobel}, {Mermilliod}, {Robichon}, {Kovalevsky}, \&
  {Crifo}}]{pbl+98}
{Perryman}, M.~A.~C., {Brown}, A.~G.~A., {Lebreton}, Y., {Gomez}, A., {Turon},
  C., {de Strobel}, G.~C., {Mermilliod}, J.~C., {Robichon}, N., {Kovalevsky},
  J., \& {Crifo}, F. 1998, \aap, 331, 81

\bibitem[{{Pinsonneault} {et~al.}(1989){Pinsonneault}, {Kawaler}, {Sofia}, \&
  {Demarque}}]{pks+89}
{Pinsonneault}, M.~H., {Kawaler}, S.~D., {Sofia}, S., \& {Demarque}, P. 1989,
  \apj, 338, 424

\bibitem[{{Pont} {et~al.}(2007){Pont}, {Gilliland}, {Moutou}, {Charbonneau},
  {Bouchy}, {Brown}, {Mayor}, {Queloz}, {Santos}, \& {Udry}}]{pgm+07}
{Pont}, F., {Gilliland}, R.~L., {Moutou}, C., {Charbonneau}, D., {Bouchy}, F.,
  {Brown}, T.~M., {Mayor}, M., {Queloz}, D., {Santos}, N., \& {Udry}, S. 2007,
  \aap, 476, 1347

\bibitem[{{Prosser} {et~al.}(1995){Prosser}, {Shetrone}, {Dasgupta}, {Backman},
  {Laaksonen}, {Baker}, {Marschall}, {Whitney}, {Kuijken}, \&
  {Stauffer}}]{psd+95}
{Prosser}, C.~F., {Shetrone}, M.~D., {Dasgupta}, A., {Backman}, D.~E.,
  {Laaksonen}, B.~D., {Baker}, S.~W., {Marschall}, L.~A., {Whitney}, B.~A.,
  {Kuijken}, K., \& {Stauffer}, J.~R. 1995, \pasp, 107, 211

\bibitem[{{Prosser} {et~al.}(1993){Prosser}, {Shetrone}, {Marilli}, {Catalano},
  {Williams}, {Backman}, {Laaksonen}, {Adige}, {Marschall}, \&
  {Stauffer}}]{psm+93}
{Prosser}, C.~F., {Shetrone}, M.~D., {Marilli}, E., {Catalano}, S., {Williams},
  S.~D., {Backman}, D.~E., {Laaksonen}, B.~D., {Adige}, V., {Marschall}, L.~A.,
  \& {Stauffer}, J.~R. 1993, \pasp, 105, 1407

\bibitem[{{Radick} {et~al.}(1987){Radick}, {Thompson}, {Lockwood}, {Duncan}, \&
  {Baggett}}]{rtl+87}
{Radick}, R.~R., {Thompson}, D.~T., {Lockwood}, G.~W., {Duncan}, D.~K., \&
  {Baggett}, W.~E. 1987, \apj, 321, 459

\bibitem[{{Sarrazine} {et~al.}(2000){Sarrazine}, {Steinhauer}, {Deliyannis},
  {Sarajedini}, {Hainline}, {Bailyn}, {Platais}, {Kozhurina-Platais}, \& {von
  Hippel}}]{ssd+00}
{Sarrazine}, A.~R., {Steinhauer}, A.~J.~B., {Deliyannis}, C.~P., {Sarajedini},
  A., {Hainline}, L.~J., {Bailyn}, C., {Platais}, I., {Kozhurina-Platais}, V.,
  \& {von Hippel}, T. 2000, in Bulletin of the American Astronomical Society,
  Vol.~32, Bulletin of the American Astronomical Society, 1461--+

\bibitem[{{Scargle}(1982)}]{scargle82}
{Scargle}, J.~D. 1982, \apj, 263, 835

\bibitem[{{Schatzman}(1962)}]{schatzman62}
{Schatzman}, E. 1962, Annales d'Astrophysique, 25, 18

\bibitem[{{Schuler} {et~al.}(2003){Schuler}, {King}, {Fischer}, {Soderblom}, \&
  {Jones}}]{skf+03}
{Schuler}, S.~C., {King}, J.~R., {Fischer}, D.~A., {Soderblom}, D.~R., \&
  {Jones}, B.~F. 2003, \aj, 125, 2085

\bibitem[{{Shu} {et~al.}(1994){Shu}, {Najita}, {Ostriker}, {Wilkin}, {Ruden},
  \& {Lizano}}]{sno+94}
{Shu}, F., {Najita}, J., {Ostriker}, E., {Wilkin}, F., {Ruden}, S., \&
  {Lizano}, S. 1994, \apj, 429, 781

\bibitem[{{Sills} {et~al.}(2000){Sills}, {Pinsonneault}, \& {Terndrup}}]{spt00}
{Sills}, A., {Pinsonneault}, M.~H., \& {Terndrup}, D.~M. 2000, \apj, 534, 335

\bibitem[{{Silva-Valio} {et~al.}(2010){Silva-Valio}, {Lanza}, {Alonso}, \&
  {Barge}}]{sla+10}
{Silva-Valio}, A., {Lanza}, A.~F., {Alonso}, R., \& {Barge}, P. 2010, \aap,
  510, A25+

\bibitem[{{Skumanich}(1972)}]{skumanich72}
{Skumanich}, A. 1972, \apj, 171, 565

\bibitem[{{Soderblom}(2010)}]{soderblom10}
{Soderblom}, D.~R. 2010, \araa, 48, 581

\bibitem[{{Soderblom} {et~al.}(2001){Soderblom}, {Jones}, \& {Fischer}}]{sjf01}
{Soderblom}, D.~R., {Jones}, B.~F., \& {Fischer}, D. 2001, \apj, 563, 334

\bibitem[{{Soderblom} {et~al.}(1983){Soderblom}, {Jones}, \& {Walker}}]{sjw83}
{Soderblom}, D.~R., {Jones}, B.~F., \& {Walker}, M.~F. 1983, \apjl, 274, L37

\bibitem[{{Soderblom} {et~al.}(1993){Soderblom}, {Stauffer}, {Hudon}, \&
  {Jones}}]{ssh+93}
{Soderblom}, D.~R., {Stauffer}, J.~R., {Hudon}, J.~D., \& {Jones}, B.~F. 1993,
  \apjs, 85, 315

\bibitem[{{Solanki} \& {Unruh}(2004)}]{su04}
{Solanki}, S.~K., \& {Unruh}, Y.~C. 2004, \mnras, 348, 307

\bibitem[{{Stassun} {et~al.}(2004){Stassun}, {Ardila}, {Barsony}, {Basri}, \&
  {Mathieu}}]{sab+04}
{Stassun}, K.~G., {Ardila}, D.~R., {Barsony}, M., {Basri}, G., \& {Mathieu},
  R.~D. 2004, \aj, 127, 3537

\bibitem[{{Stassun} {et~al.}(1999){Stassun}, {Mathieu}, {Mazeh}, \&
  {Vrba}}]{smm+99}
{Stassun}, K.~G., {Mathieu}, R.~D., {Mazeh}, T., \& {Vrba}, F.~J. 1999, \aj,
  117, 2941

\bibitem[{{Stassun} {et~al.}(2001){Stassun}, {Mathieu}, {Vrba}, {Mazeh}, \&
  {Henden}}]{smv+01}
{Stassun}, K.~G., {Mathieu}, R.~D., {Vrba}, F.~J., {Mazeh}, T., \& {Henden}, A.
  2001, \aj, 121, 1003

\bibitem[{{Stassun} {et~al.}(2007){Stassun}, {van den Berg}, \&
  {Feigelson}}]{svf07}
{Stassun}, K.~G., {van den Berg}, M., \& {Feigelson}, E. 2007, \apj, 660, 704

\bibitem[{{Stauffer} {et~al.}(1984){Stauffer}, {Hartmann}, {Soderblom}, \&
  {Burnham}}]{shs+84}
{Stauffer}, J.~R., {Hartmann}, L., {Soderblom}, D.~R., \& {Burnham}, N. 1984,
  \apj, 280, 202

\bibitem[{{Stauffer} \& {Hartmann}(1987)}]{sh87}
{Stauffer}, J.~R., \& {Hartmann}, L.~W. 1987, \apj, 318, 337

\bibitem[{{Stauffer} {et~al.}(1985){Stauffer}, {Hartmann}, {Burnham}, \&
  {Jones}}]{shb+85}
{Stauffer}, J.~R., {Hartmann}, L.~W., {Burnham}, J.~N., \& {Jones}, B.~F. 1985,
  \apj, 289, 247

\bibitem[{{Stauffer} {et~al.}(1989){Stauffer}, {Hartmann}, \& {Jones}}]{shj89}
{Stauffer}, J.~R., {Hartmann}, L.~W., \& {Jones}, B.~F. 1989, \apj, 346, 160

\bibitem[{{Steinhauer}(2010)}]{steinhauer10}
{Steinhauer}, A. 2010, priv. comm.

\bibitem[{{Sung} \& {Bessell}(1999)}]{sb99}
{Sung}, H., \& {Bessell}, M.~S. 1999, \mnras, 306, 361

\bibitem[{{Terndrup} {et~al.}(2002){Terndrup}, {Pinsonneault}, {Jeffries},
  {Ford}, {Stauffer}, \& {Sills}}]{tpj+02}
{Terndrup}, D.~M., {Pinsonneault}, M., {Jeffries}, R.~D., {Ford}, A.,
  {Stauffer}, J.~R., \& {Sills}, A. 2002, \apj, 576, 950

\bibitem[{{Terndrup} {et~al.}(2000){Terndrup}, {Stauffer}, {Pinsonneault},
  {Sills}, {Yuan}, {Jones}, {Fischer}, \& {Krishnamurthi}}]{tsp+00}
{Terndrup}, D.~M., {Stauffer}, J.~R., {Pinsonneault}, M.~H., {Sills}, A.,
  {Yuan}, Y., {Jones}, B.~F., {Fischer}, D., \& {Krishnamurthi}, A. 2000, \aj,
  119, 1303

\bibitem[{{Tinker} {et~al.}(2002){Tinker}, {Pinsonneault}, \&
  {Terndrup}}]{tpt02}
{Tinker}, J., {Pinsonneault}, M., \& {Terndrup}, D. 2002, \apj, 564, 877

\bibitem[{{Vasilevskis} {et~al.}(1958){Vasilevskis}, {Klemola}, \&
  {Preston}}]{vkp58}
{Vasilevskis}, S., {Klemola}, A., \& {Preston}, G. 1958, \aj, 63, 387

\bibitem[{{Weber} \& {Davis}(1967)}]{wd67}
{Weber}, E.~J., \& {Davis}, Jr., L. 1967, \apj, 148, 217

\bibitem[{{Yi} {et~al.}(2003){Yi}, {Kim}, \& {Demarque}}]{ykd03}
{Yi}, S.~K., {Kim}, Y., \& {Demarque}, P. 2003, \apjs, 144, 259

\end{thebibliography}


\end{document}